\documentclass{ws-ijmpa}
\usepackage[super,compress]{cite}
\usepackage{graphicx}

\usepackage[english]{babel}
\usepackage{t1enc}
\usepackage[utf8]{inputenc}
\usepackage[T1]{fontenc}
\usepackage{amsmath,amssymb,amsfonts}
\usepackage{indentfirst}
\usepackage{color}
\usepackage{array}
\usepackage{verbatim}
\usepackage{comment}

\newcommand{\nc}{\newcommand}
\nc{\td}{\mathrm{d}}
\nc{\kd}{\mathbf{k}}
\nc{\Kd}{\mathbf{K}}
\nc{\qd}{\mathbf{q}}
\nc{\Rd}{\mathbf{R}}
\nc{\rd}{\mathbf{r}}
\nc{\ud}{\mathbf{u}}
\nc{\vd}{\mathbf{v}}
\nc{\pd}{\mathbf{p}}
\nc{\Pd}{\mathbf{P}}
\nc{\Gd}{\mathbf{G}}
\nc{\Ad}{\mathbf{A}}
\nc{\Dd}{\mathbf{D}}
\nc{\Sd}{\mathbf{S}}
\nc{\md}{\mathbf{m}}
\nc{\Fd}{\mbox{\boldmath $F$}}
\nc{\Md}{\mbox{$\mathcal{M}$}}
\nc{\bd}{\mbox{\boldmath $\beta$}}
\nc{\Od}{\mbox{$\mathbf{\mathcal{O}}$}}
\nc{\od}{\mbox{\boldmath $\omega$}}
\nc{\Ddd}{\mbox{\boldmath $\underline{\underline{D}}$}}
\nc{\Mdd}{\mbox{\boldmath $\underline{\underline{M}}$}}
\nc{\Cd}{\mbox{\boldmath $\underline{\underline{C}}$}}
\nc{\Div}{\mathrm{div}}
\nc{\Rot}{\mathrm{rot}}
\nc{\Grad}{\mathrm{grad}}
\nc{\Det}{\mathrm{det}}

\let\originalleft\left 
\let\originalright\right 
\renewcommand{\left}{\mathopen{}\mathclose\bgroup\originalleft} 
\renewcommand{\right}{\aftergroup\egroup\originalright}

\begin{document}
	\markboth{G.~Kasza, T.~Cs\"org\H{o}}{Lifetime estimations from RHIC Au+Au data}
	
	%
	\catchline{}{}{}{}{}
	%
	
	\title{Lifetime estimations from RHIC Au+Au data}
	
	\author{G\'abor Kasza}
	
	\address{
		Wigner RCP, H-1525 Budapest 114, P.O.Box 49, Hungary\\
		EKU KRC, H-3200 Gy\"ongy\"os, M\'atrai \'ut 36, Hungary\\ 
		E\"otv\"os University, H-1117 Budapest, P\'azm\'any P. s. 1/A, Hungary\\
		kasza.gabor@wigner.mta.hu}
	
	\author{Tam\'as Cs\"org\H{o}}
	
	\address{
		Wigner RCP, H-1525 Budapest 114, P.O.Box 49, Hungary\\
		EKU KRC, H-3200 Gy\"ongy\"os, M\'atrai \'ut 36, Hungary\\
		CERN, CH-1211 Geneve 23, Switzerland\\
		tcsorgo@cern.ch}
	
	\maketitle
	
	\begin{history}
		\received{Day Month Year}
		\revised{Day Month Year}
	\end{history}
	
	\begin{abstract}
		We discuss a recently found family of exact and analytic, finite and accelerating, 1+1 dimensional solutions of perfect fluid relativistic hydrodynamics to describe the pseudorapidity densities and longitudinal HBT-radii and to estimate the lifetime parameter and the initial energy density of the expanding fireball in Au+Au collisions at RHIC with $\sqrt{s_{NN}}=130$ GeV and $200$ GeV  colliding energies. From these exact solutions of relativistic hydrodynamics, we derive a simple and powerful formula to describe the pseudorapidity density distributions in high energy proton-proton and heavy ion collisions, and derive the scaling of the longitudinal HBT radius parameter as a function of  the pseudorapidity density. We improve upon several oversimplifications in Bjorken's famous initial energy density estimate, and  apply our results to estimate the initial energy densities of high energy reactions with data-driven pseudorapidity distributions. When compared to similar estimates at the LHC energies, our results indicate a surprising and non-monotonic dependence of the initial energy density on the energy of heavy ion collisions.
		
		\keywords{relativistic hydrodynamics; quark-gluon plasma; longitudinal flow; rapidity distribution; pseudorapidity distribution; HBT-radii; initial energy density.}
	\end{abstract}
	
	\ccode{PACS numbers:}
	
	\section{Introduction}
	Relativistic hydrodynamics of nearly perfect fluids is the current paradigm in analyzing soft particle production processes in high energy heavy ion collisions. 
	The development of this paradigm
	goes back to the classic papers of Fermi from 1950~\cite{Fermi:1950jd}, Landau from 1953~\cite{Landau:1953gs} and Bjorken from 1982\cite{Bjorken:1982qr},
	that analyzed the statistical and collective aspects of multiparticle production in high energy collisions of elementary particles and atomic nuclei.
	
	Although hydrodynamical relations in the double-differential invariant momentum distribution and in the parameters of the Bose-Einstein correlation functions of hadron-proton collisions were observed at $\sqrt{s} = 22 $ GeV colliding energies as early as in 1998~\cite{Agababyan:1997wd}, relativistic hydrodynamics of nearly perfect fluids became the dominant paradigm for heavy ion collisions only after 2004, following the publication of the White Papers of the four RHIC experiments in refs. 
	BRAHMS~\cite{Arsene:2004fa}, 
	PHENIX~\cite{Adcox:2004mh}, 
	PHOBOS~\cite{Back:2004je} and 
	STAR~\cite{Adams:2005dq}. 
	These papers summarized the results of the first four years of data-taking at Brookhaven National Laboratory's Relativistic Heavy Ion Collider (BNL's RHIC) as a circumstantial evidence for creating a nearly perfect fluid of strongly coupled quark-gluon plasma or quark matter in $\sqrt{s_{NN}} = 200$ GeV Au+Au collisions.
	
	Recently, the STAR Collaboration significantly sharpened this result by pointing out that the fluid created in heavy ion collisions at RHIC is not only the most perfect fluid ever made by humans but also the most vortical fluid as well~\cite{STAR:2017ckg}.
	Importantly, the PHENIX collaboration demonstrated that the domain of validity of the hydrodynamical paradigm extends to p + Au, d + Au and $^3$He + Au collisions~\cite{PHENIX:2018lia}
	and also to as low energies as $\sqrt{s_{NN}} = 19.6$ GeV~\cite{Aidala:2017ajz}.
	So by 2018, the dominant paradigm of
	analyzing the collisions of hadron-nucleus and small-large nucleus collisions also shifted to
	the domain of relativistic hydrodynamics, as reviewed recently by Ref.~\citen{Nagle:2018nvi}.
	Results by the ALICE Collaboration at LHC suggested that enhanced production of multi-strange
	hadrons in high-multiplicity proton-proton collisions may signal the production of a strongly coupled
	quark-gluon plasma not only in high energy proton/deuteron/Helium+ nucleus but also in hadron-hadron collisions
	~\cite{ALICE:2017jyt}. These results are re-opening the door to the application of the tools of relativistic hydrodynamics in hadron-hadron collisions too, however the applicability of the hydrodynamical paradigm
	in these collisions is currently under intense theoretical and experimental scrutinity. Although more than 20 years passed since hydrodynamical couplings were found in the double-differential invariant momentum distribution and in the parameters of the Bose-Einstein correlation functions of hadron-proton collisions~\cite{Agababyan:1997wd}, this field is still a subject of intense debate, where the picture that protons at high colliding energies behave in several ways similarly to a small nucleus~\cite{Nemes:2015iia}
	are gaining popularity but not yet well known.  
	
	The theory and applications of relativistic hydrodynamics to high energy collisions was reviewed and detailed recently in Ref.~\citen{deSouza:2015ena}, with special attention also to the role of analytic solutions, but with a primary interest in the hydrodynamical interpretation of azimuthal oscillations in the transverse  momentum spectra, and its relation to the  transverse hydrodynamical expansion. 
	In contrast, the focus of our manuscript is to gain a deeper understanding of the longitudinal expansion dynamics of high energy heavy ion and proton-proton collisions. As reviewed recently in Ref.~\citen{Csorgo:2018pxh}, the longitudinal momentum spectra of high energy collisions can a posteriori be very well described by exact solutions of 1+1 dimensional solutions of perfect fluid hydrodynamics, the central theme of our current investigation.

	Let us also stress that this manuscript is the fourth part of a manuscript series, that is a
	straigthforward continuation of investigations published in Refs.~\citen{Csorgo:2018fbz,Kasza:2018jtu,Csorgo:2018crb}.
	The first part of this series~\cite{Csorgo:2018fbz} gives an introductory overview of the special research area of 1+1 dimensional exact solutions of relativistic hydrodynamics, defines the notation that we also utilize in the current work,  and summarizes a recently found new  class of exact solutions of perfect fluid relativistic hydrodynamics in 1+1 dimensions.
	This new family of exact solutions, the CKCJ solution was discovered by Cs\"org\H{o}, Kasza, Csan\'ad and Jiang and presented first in Ref.~\citen{Csorgo:2018pxh}.
	
	The first part of this manuscript series~\cite{Csorgo:2018fbz} connects the CKCJ solution to  experimentally measured quantities by  evaluating the rapidity and pseudorapidity density distributions. The second part of this manuscript series~\cite{Kasza:2018jtu} presents  exact results for the initial energy density in high energy collisions that at the same time represent an apparently  fundamental correction to the famous Bjorken estimate of initial energy density~\cite{Bjorken:1982qr} in high energy collisions. 
	The third part of this  manuscript series~\cite{Csorgo:2018crb}, evaluates the Bose-Einstein correlation functions and in a Gaussian approximation it determines  the  so-called Hanbury Brown - Twiss or HBT-radii in the longitudinal direction from the CKCJ solution in order to determine the life-time of the fireball created in high-energy collisions. 
	
	In this work, corresponding to the  fourth part of this manuscript series, we started to investigate the excitation function of the parameters of the initial state, as reconstructed with the help of the CKCJ solution ~\cite{Csorgo:2018pxh}. The results demonstrate the advantage of analytic solutions in understanding the longitudinal dynamics of fireball evolution in high energy heavy ion collisions. We follow up Refs.~\citen{Csorgo:2018fbz,Kasza:2018jtu,Csorgo:2018crb} by presenting  a new method
	to evaluate the lifetime and the initial energy density of high energy heavy-ion collisions, utilizing the recently found CKCJ family of solutions~\cite{Csorgo:2018fbz,Kasza:2018jtu,Csorgo:2018crb} to describe their longitudinal expansion and the corresponding measurable, the pseudorapidity density distributions.
	Namely, in this paper we show new fit results for Au+Au at $\sqrt{s_{NN}}=130$ GeV and Au+Au at $\sqrt{s_{NN}}=200$ GeV collisions at RHIC. Through these calculations we are able to estimate the acceleration and the effective temperature of the medium at freeze-out. In this paper, we utilize these results to determine the lifetime of the fireball by fitting the longitudinal HBT-radii data of Au+Au at $\sqrt{s_{NN}}=130$ GeV and Au+Au at $\sqrt{s_{NN}}=200$ GeV collisions, simultaneously $dN/d\eta_p$ and the slope of the transverse momentum spectra are also described. In Ref.~\citen{Kasza:2018jtu} we presented a new exact formula of the initial energy density derived from our new family of perfect fluid hydrodynamic solutions. In this new formula all the unknown parameters
	of the freeze-out stage can be determined by fits to pseudorapidity densities and longitudinal radii, however, the dependence on  the initial proper time $\tau_0$ remains explicit. As a consequence, our new method makes it possible to describe the initial proper time dependence of the initial energy density of the thermalized fireball in high-energy heavy-ion collisions, for predominantly 1+1 dimensional expansions.
	
	\section{New, exact solutions of perfect fluid hydrodynamics}
	The equations of relativistic perfect fluid hydrodynamics express the local conservation of entropy and four-momentum:
	\begin{eqnarray}
	\partial_{\mu}\left(\sigma u^{\mu}\right)&=&0, 
	\label{e:entropy} \\
	\partial_{\nu}T^{\mu \nu} &= &0, 
	\label{e:energy-momentum} 
	\end{eqnarray}
	where $\sigma$ is the entropy density, the four velocity is denoted by $u^{\mu}$ and normalized as $u^{\mu}u_{\mu}=1$, and $T^{\mu \nu}$ stands for the energy-momentum four tensor of perfect fluids:
	\begin{equation} 
	T^{\mu \nu}=\left(\varepsilon+p\right)u^{\mu}u^{\nu} - pg^{\mu \nu}.
	\end{equation}
	The metric tensor is $g^{\mu
		\nu}=\textnormal{diag}(1,-1,-1,-1)$, the pressure is denoted by $p$, and the energy density by $\varepsilon$. The entropy density $\sigma\equiv \sigma(x)$, the energy density $\varepsilon\equiv \varepsilon(x)$, the temperature $T\equiv T(x)$, the pressure $p \equiv p(x)$, the four-velocity $u^\mu\equiv u^{\mu}(x) $ and the four-momentum tensor $T^{\mu \nu}\equiv T^{\mu \nu}(x) $ are fields, i.e. they are functions of the four coordinate $x \equiv x^{\mu}=(t,\mathbf{r})=\left(t,r_x,r_y,r_z\right)$, but these dependencies are suppressed in our notation. The above set of equations provides five equations for six unknown fields, and it is closed by an equation of state (EoS). 
	For a broad class of equations of state,  we may assume that the energy density $\varepsilon$ is proportional to the pressure $p$ with a temperature dependent proportionality factor $\kappa(T)$:
	\begin{equation}
	\varepsilon=\kappa(T) p.
	\end{equation}
	This class of equations of state was shown to be thermodynamically consistent
	in Refs.~\citen{Csorgo:2001xm,Csanad:2012hr}.
	For the case of zero baryochemical potential,  this equation of state implies a temperature dependent function for the speed of sound $	c_s $, that can be expressed in terms of $\kappa$ as follows:
	\begin{equation}
	c_s(T) = \sqrt{\frac{\partial p}{\partial \varepsilon}} =  \frac{1}{\sqrt{\kappa(T)}}.
	\end{equation}
	As shown in Ref.~\citen{Csanad:2012hr}, the lattice QCD equation of state at $\mu_B = 0$ belongs to this class
	and leads to exact solutions of fireball hydrodynamics. For more details of exact hydrodynamical solutions
	that use the temperature dependent speed of sound from lattice QCD, for the simplest examples and details we may refer to
	Ref.~\citen{Csorgo:2001xm,Csanad:2012hr} and their generalizations, including Refs.~\citen{Csorgo:2016ypf,Csorgo:2013ksa}
	where one can clearly see on the level of equations that a temperature independent, average speed of sound may correspond
	on the average to similar exact solutions. This feature is also demonstrated on Fig. 2 of Ref.~\citen{Csanad:2012hr} where it is shown that the time evolution of the temperature in the center of a three-dimensionally expanding relativistic fireball
	for the lattice QCD equation of state may be actually rather close to a that of a similarly three-dimensionally expanding fireball
	but for an average value of a temperature independent speed of sound, corresponding to $\kappa \approx 4$, corresponding to
	$\langle c_s(T)\rangle \approx 0.5$.
	We conjecture that the temperature (in)dependence of the speed of sound is not a fundamentally difficult and analytically
	inaccessible feature of exact solutions of relativistic fireball hydrodynamics. However, for the sake of simplicity
	and in order to be able to evaluate the experimental consequences of the CKCJ exact solution of relativistic, 1+1 dimensional
	hydrodynamics,   we utilize a temperature independent,
	average value for the speed of sound from now on. This step corresponds to the $\kappa (T) = \kappa$ approximation. As our goal is to describe experimental data, we take the average value of the speed of sound from a PHENIX measurement~\cite{Adare:2006ti}, 
	that corresponds to $\langle c_s(T)\rangle = 1/\sqrt{\kappa} = 0.35 \pm 0.05 $. This PHENIX value of the average speed of sound corresponds  to $\kappa \approx  10 {+1 \atop{-3}}$. The result of this PHENIX measurements thus seems to be  different from the average value of the speed of sound from Monte Carlo lattice QCD simulations~\cite{Borsanyi:2010cj}, that corresponds to $\kappa_{\rm lQCD} \approx 4 \pm 1 $ according  to Fig. 2 of Ref.~\citen{Csanad:2012hr}.
	
	\subsection{The CKCJ solution}
	An exact and analytic, finite and accelerating, 1+1 dimensional solution of relativistic perfect fluid hydrodynamics was recently found by Cs\"org\H{o}, Kasza, Csan\'ad and Jiang (CKCJ)~\cite{Csorgo:2018pxh} 
	as a family of parametric curves. The thermodynamic parameters as the entropy density $\sigma$ and the temperature $T$ are functions of the longitudinal proper time $\tau$ and the space-time rapidity $\eta_x$:
	\begin{equation}
	\left(\tau,\eta_x\right) =  
	\left(\, \sqrt{t^2-r_z^2}\, ,           
	\frac{1}{2}\textnormal{ln}\left[\frac{t+r_z}{t-r_z}\right]\,\right).
	\end{equation}
	The fluid rapidity $\Omega$ is assumed to be independent of the proper time, $\Omega(\tau,\eta_x) \equiv \Omega(\eta_x)$. The four-velocity is chosen as $u^{\mu}= \left(\cosh\left(\Omega\right),\sinh\left(\Omega\right)\right)$, consequently the three-velocity is $v_z=\tanh\left(\Omega\right)$. The new class of CKCJ solutions~\cite{Csorgo:2018pxh} is given in terms of parametric curves, that can be summarized as follows:
	\begin{eqnarray}
	\eta_x(H)  & = & \Omega(H) -H, 
	\label{e:etaH}\\ 
	\Omega(H)  & = & 
	\frac{\lambda}{\sqrt{\lambda-1}\sqrt{\kappa-\lambda}}
	\textnormal{arctan}\left(\sqrt{\frac{\kappa-\lambda}
		{\lambda-1}}\textnormal{tanh}\left(H\right)\right), 
	\label{e:OmegaH} \\ 
	\sigma(\tau,H)&= & \sigma_0 
	\left(\frac{\tau_0}{\tau}\right)^{\lambda}
	\mathcal{V}_{\sigma}(s) \left[1+\frac{\kappa-1}{\lambda-1}
	\textnormal{sinh}^2(H)\right]^{-\frac{\lambda}{2}},
	\label{e:sigmasol} \\
	T(\tau,H)  & = & T_0 
	\left(\frac{\tau_0}{\tau}\right)^{\frac{\lambda}{\kappa}} 
	\mathcal{T}(s) 
	\left[1+\frac{\kappa-1}{\lambda-1}\textnormal{sinh}^2(H)
	\right]^{-\frac{\lambda}{2\kappa}},
	\label{e:Tsol}\\ 
	\mathcal{T}(s) & = & 
	\frac{1}{\mathcal{V}_{\sigma}(s)},
	\label{e:scalingsol}\\
	s(\tau,H) & = & 
	\left(\frac{\tau_0}{\tau}\right)^{\lambda-1} 
	\textnormal{sinh}(H)\left[1 + \frac{\kappa-1}{\lambda-1}
	\textnormal{sinh}^2(H)\right]^{-\lambda/2}.
	\label{e:sH}
	\end{eqnarray}
	Here $H$ is the parameter of the parametric curves that specify the CKCJ solution, $\lambda$ is the acceleration parameter, $s(\tau,H)$ stands for the scale variable, and $\mathcal{V}_{\sigma}(s)$ is an arbitrary positive definite scaling function for the entropy density. Physically, $H $ is the difference of the fluid rapidity and the coordinate space rapidity. The integration constants $\sigma_0$ and $T_0$ stand for $\sigma(\tau_0,H=0)$ and $T(\tau_0,H=0)$, where the initial proper time is denoted by $\tau_0$. The space-time rapidity dependence appears only through the parameter $H$, which is the difference of the fluid rapidity $\Omega$ and the coordinate rapidity $\eta_x$. The CKCJ solutions are limited to a cone within the forward light-cone around midrapidity. The domain of validity of these solutions in  space-time
	rapidity is described in details in Ref.~\citen{Csorgo:2018pxh}.

	\section{Observables}
	This section presents the results of the observables that are derived from the CKCJ solution. First we start with the derivation of the mean multiplicity. This part is detailed here as this was not discussed in the literature before. Then we discuss the pseudorapidity density and the longitudinal HBT-radii, that are the key longitudinal dynamics dependent observables to determine the lifetime parameter of the fireball. In this manuscript we do not detail the derivation of these quantities, as these were given in Refs.~\citen{Kasza:2018jtu, Csorgo:2018crb} before. In the last subsection of this section, we derive a scaling relation
	between the longitudinal HBT radius parameter and the pseudorapidity density.
	
	\subsection{Mean multiplicity}
	\label{ss:mean-n}
	In order to obtain the mean multiplicity $\langle N  \rangle$ and 
	the  pseudorapidity density $\frac{dN}{d\eta_p}$, we follow Refs.~\citen{Kasza:2018jtu, Csorgo:2018crb} and
	embed these 1+1 dimensional family of solutions to the 1+3 dimensional space-time. For the sake of simplicity and clarity, we also assume, that the freeze-out hypersurface is pseudo-orthogonal to the four velocity. 
	
	In this case, the average multiplicity $\langle N\rangle$ can be calculated from the phase-space distribution with integrals over the space-time and momentum-space volumes,
	\begin{equation}
	\langle N \rangle = \frac{g}{(2\pi \hbar)^3} \int d^4 x \int \frac{d^3p}{p^0} p^{\mu} \Sigma_{\mu}(x)\exp\left(-\frac{p_{\mu}u^{\mu}}{T(x)}+\frac{\mu(x)}{T(x)}\right),
	\end{equation}
	where $g=2s+1$ is the spin degeneration factor, 
	the chemical potential is the function of the space-time coordinates and denoted by $\mu(x) \equiv \mu$, 
	and $\Sigma_{\mu}(x) \equiv \Sigma_{\mu}$ stands for the normal vector of the freeze-out hypersurface.
	As we have shown in Ref.~\citen{Csorgo:2018pxh} this implies for the freeze-out hypersurface of the CKCJ solution,
	that
	\begin{equation}
	\Sigma_{\mu}=u_{\mu}\:\frac{\delta (\tau-\tau_f)}{\cosh\left(\Omega-\eta_x\right)}.
	\end{equation}
	Thus the invariant integration over the momentum  can be performed at each point on the freeze-out hypersurface,
	parameterized as $\tau=\tau(\eta_x)$, to find
	\begin{equation}\label{e:average_N2}
	\langle N \rangle = \frac{g}{(2\pi \hbar)^3} 4\pi m^2 \int d\tau\: d\eta_x\: dr_x\: dr_y \: \delta (\tau-\tau_f)\frac{\tau T }{\cosh\left(\Omega-\eta_x\right)} K_2\left(\frac{m}{T}\right)\exp\left(\frac{\mu}{T}\right),
	\end{equation}
	where $K_2$ stands for the modified Bessel-function of the second kind and the integral 
	measure is rewritten as $d^4x = \tau\:d\tau\:d\eta_x\: dr_x\:dr_y$.
	We assume, that close to  the midrapidity region the temperature is approximately independent from $\eta_x$ and the transverse coordinates so it is approximately
	a constant freeze-out temperature $T_f$. Using these approximations,
	Eq.~\eqref{e:average_N2} can be simplified as:
	\begin{equation}
	\langle N \rangle = \frac{g}{(2\pi \hbar)^3} 4\pi m^2 T_f K_2\left(\frac{m}{T_f}\right)\int d\eta_x\: dr_x\: dr_y \:\frac{\tau_f}{\cosh\left(\Omega-\eta_x\right)} \exp\left(\frac{\mu}{T_f}\right), \label{e:mean-multiplicity}
	\end{equation}
	where we note the suppressed coordinate-space rapidity $\eta_x$ and transverse coordinate $(r_x,r_y)$ dependence of the fugacity factor $\exp(\frac{\mu}{T_f})$. As  already clear
	from Ref.~\citen{Csorgo:2018pxh},  this fugacity factor contributes importantly to the shape of coordinate rapidity density and
	hence to the measurable (pseudo)rapidity distribution. At the end of this sub-section, we shall approximate this factor by Gaussians both in $\eta_x$ and in the transverse coordinates $(r_x,r_y)$.
	
	Let us also introduce $V_x$ for the spatial volume, and $V_p$ for the local volume of the momentum space (corresponding to particles emitted at the same coordinate position $x$.) Let us also denote by $\bar{f}$  the average phase space density. 
	
	The average number of particles $\langle N\rangle$ -- also referred to as the mean multiplicity --   is given by these quantities in a rather trivial and straight-forward manner:
	\begin{equation}\label{e:average_N}
	\langle N \rangle = \frac{g}{\left(2\pi \hbar \right)^3}\bar{f}\:V_x V_p,
	\end{equation}
	where the average phase-space density is $\bar{f}=\exp(\bar{\mu}/T_f)$  and $\bar{\mu}$ stands for the average value of
	the chemical potential for the particles characterized by mass $m$.
	From Eqs.~\eqref{e:mean-multiplicity} and \eqref{e:average_N} one finds the formulae of the volumes of both the coordinate-space 
	($V_x$ given in terms of the fugacity distribution) and  of the momentum space ($V_p$, given in terms of the freeze-out 
	temperature $T_f$ and the mass $m$) as:
	\begin{align}
	V_x&=\int d\eta_x\: dr_x\: dr_y \:\frac{\tau_f}{\cosh\left(\Omega-\eta_x\right)} \exp\left(\frac{\mu(\tau_f,\eta_x,r_x,r_y)-\bar{\mu}-m}{T_f}\right),\label{e:Vx}\\
	V_p&=4\pi m^2 T_f K_2\left(\frac{m}{T_f}\right)\exp\left(\frac{m}{T_f}\right).
	\end{align}
	When we implement a Gaussian approximation to perform the integrals in Eq.~\eqref{e:Vx} we obtained:
	\begin{equation}
	V_x \approx \left(2\pi R_G^2\right)\left(2\pi \tau_f^2 \Delta \eta_x^2\right)^{1/2},
	\end{equation}
	where $R_G$ is the Gaussian width of the transverse coordinate distribution and $\Delta \eta_x$ is the same for the space-time rapidity distribution. In the non-relativistic limit, the  volume of the local momentum-space distribution can be simplified as:
	\begin{equation}\label{e:Vp_limit}
	V_p\approx \left(2\pi m T_f\right)^{3/2}.
	\end{equation}
	These leading order Gaussian limiting cases and approximations will be  discussed and elaborated further in the sub-section on the discussion of the scaling properties of the longitudinal HBT-radii $R_{\rm long}$, in subsection~\ref{ss:Rlong}. For now, the key point of this sub-section was that in order to evaluate the mean multiplicity, we can perform the invariant integrals over the momentum variables first, going to the local rest frame at each point on the freeze-out hypersurface if on these hypersurfaces
	the temperature has an approximately  constant value of $T_f$. This way a J\"uttner pre-factor can be pulled out with the average value of the fugacity and the remaining integral will be the average phase-space density multiplied by the total freeze-out volume (in coordinate-space) of the fireball.

	\subsection{Pseudorapidity density}
	\label{ss:dndeta}
	As detailed in Refs.~\citen{Csorgo:2018pxh, Csorgo:2018fbz} the analytic expression of the rapidity distribution was calculated by an advanced saddle-point integration to yield
	\begin{equation}
	\frac{dN}{dy} \approx
	\left.\frac{dN}{dy}\right|_{y=0} 
	\cosh^{-\frac{1}{2}\alpha(\kappa,\lambda)-1}\left(\frac{y}{\alpha(1,\lambda)}\right)
	\exp\left(-\frac{m}{T_{\rm eff}} 
	\left[\cosh^{\alpha(\kappa,\lambda)}\left(\frac{y}{\alpha(1,\lambda)}\right)-1\right]\right).\label{e:dndy-function}
	\end{equation}
	The explicit expression for the normalization constant $	\left.\frac{dN}{dy}\right|_{y=0} $
	as a function of fit parameters is given by Eq. (36) of Ref.~\citen{Csorgo:2018pxh}. From now on,
	we recommend to use the rapidity density at mid-rapidity as a  normalization parameter fitted to data, or, as a  directly measured quantity.
	
	In this equation $y = \frac{1}{2} \ln\left(\frac{E + p_z}{E-p_z}\right)$ stands for the rapidity, the four-momentum is defined as $p^{\mu}=\left(E,p_x,p_y,p_z\right)$ with $E= \sqrt{m^2 + p^2}$,
	where the modulus of the three-momentum is $p = \sqrt{p_x^2 + p_y^2 + p_z^2}$ and $m$ stands for the particle mass. The slope parameter of the transverse momentum spectrum is denoted as $T_{\rm eff}$, standing for an effective temperature, and in order to simplify the notation we also introduce the auxiliary function
	$\alpha$ defined as
	\begin{equation}
	\alpha(\kappa,\lambda)=\frac{2\lambda-\kappa}{\lambda-\kappa}.
	\end{equation}
	Note that in our earlier papers, the dependence of this function $\alpha$ on one of its variables, $\lambda$ was suppressed, using effectively an $\alpha \equiv \alpha(\kappa,\lambda) \equiv \alpha(\kappa) $ notation in our earlier publications. 
	From now on,  we use a new, more explicit notation of $\alpha(\kappa,\lambda) $,
	to denote the same function but also to make the $\lambda$ dependence of our
	results more transparent.
	
	It is worthwhile to mention, that Eq.~\eqref{e:dndy-function} can be further approximated in the mid-rapidity region, provided that
	$|y| \ll \alpha(1,\lambda) = \frac{2 \lambda -1}{\lambda -1} = 2 + \frac{1}{\lambda -1}.$ Obviously the domain in rapidity
	where such a nearly Gaussian approximation is valid, is increasing quickly as the shape parameter 
	approaches the boost-invariant $\lambda \rightarrow 1$ limit. In such a leading order Gaussian approximation,
	the  rapidity distribution reads as
	\begin{eqnarray}
	\frac{dN}{dy}  & \approx &
	\frac{\langle N\rangle}{\big(2\pi\Delta^2 y\big)^{1/2}}
	\exp\Bigg(-\frac{y^2}{ 2 \Delta^2 y \phantom{\Big|} }  \Bigg)  \, .\label{e:dndy-Gauss}
	\end{eqnarray}
	In this Gaussian approximation, the midrapidity density and the Gaussian width of the rapidity distribution 
	can be expressed with the parameters of the CKCJ solution embedded to 1+3 dimensions  as follows:
	\begin{eqnarray}
	\left.\frac{dN}{dy}\right|_{y=0} & = & 
	\frac{	\langle N\rangle }{(2 \pi \Delta^2 y)^{\frac{1}{2}} }	, \label{e:dndy-Gauss-norm} \\
	\frac{1}{\Delta^2 y} & = & (\lambda - 1)^2 \, \left[ 1 + \left( 1 - \frac{1}{\kappa} \right) 
	\left(\frac{1}{2} + \frac{m}{T_{\rm eff}} \right) \right] .
	\label{e:dndy-Gauss-width}
	\end{eqnarray}
	where the mean multiplicity is given by Eq.~\eqref{e:mean-multiplicity}, also as a function of the parameters
	of the embedded exact solutions is given by  Eq. (36) of Ref.~\citen{Csorgo:2018pxh}, with the replacement of $T_f \rightarrow T_{\rm eff}$ in that equation. As mentioned before, in this paper we prefer  to carry on the midrapidity density or the mean multiplicity
	as one of our fit parameters.

	In this approximation, one may  use the mean multiplicity $\langle N \rangle$ and the width of the
	rapidity distribution, $\Delta y$, as the relevant fit parameters, instead of the original four fit parameters
	of the CKCJ solution of relativistic hydrodynamics.
	
	This result is a beautiful example of hydrodynamical scaling behaviour: although the CKCJ exact solution
	at the time of the freeze-out predicts that the observables will depend on four different parameters,
	$\left.\frac{dN}{dy}\right|_{y=0} $, 
	$\kappa$, $\lambda$, and $T_{\rm eff}$, actually in the $\lambda \rightarrow 1$ limiting case a 
	scaling behaviour is found: the rapidity density becomes a function of only two combinations
	of its four physical fit parameters. So different physical fit parameters may result in the same rapidity
	density, if the two relevant combinations of the fit parameters remain the same.

	From this shape it is also clear that the rapidity densities can be normalized both to the mid-rapidity density as well
	as to the total multiplicity. For an nearly boost-invariant fireball, 
	a nearly boost-invariant distribution is obtained, and in this case  its  normalization 
	to the mid-rapidity density  remains valid in the boost-invariant, $\lambda \rightarrow 1$ limit too, although
	in this limit both the mean multiplicity and the width parameter of the rapidity density diverge.
	
	By a saddle-point integration of the double differential spectra~\cite{Csorgo:2018pxh, Csorgo:2018fbz},
	the pseudorapidity density can also be calculated as a parametric curve, where the parameter is the rapidity $y$:
	\begin{equation}
	\left(\eta_p(y) \, , \frac{dN}{d\eta_p}(y)  \right) =
	\left( \frac{1}{2}\log\left[\frac{\bar{p}(y) + \bar{p}_z(y)}{\bar{p}(y)-\bar{p}_z(y)}\right] \, , 
	\frac{\bar{p}(y)}{\strut \bar{E}(y)}\frac{dN}{dy} \right).
	\label{e:dndeta}
	\end{equation}
	The pseudorapidity is denoted by $\eta_p = \frac{1}{2} \ln\left(\frac{p + p_z}{p-p_z}\right)$. The average of the momentum space  variable  $q$ is indicated by $\bar{q}$, as given in  details in Ref.~\citen{Csorgo:2018fbz}.
	In order to obtain the pseudorapidity density as an analytic function of $\eta_p$, we calculated the rapidity dependent average transverse mass of the particles:
	\begin{equation}
	\bar{m}_T(y)=\int\limits_{m}^{\infty} m_T \left(\frac{dN}{dy}\right)^{-1}\frac{dN}{dm_T\:dy}\:dm_T.
	\end{equation}
	To perform the integral we used saddle-point integration method which led us to the following function:
	\begin{equation}
	\bar{m}_T(y) \simeq m + \frac{T_{\rm eff}}{\cosh^{\alpha(\kappa,\lambda)}\left(\frac{y}{\alpha(1,\lambda)}\right)}. \label{e:mtbar-y}
	\end{equation}
	From the above \eqref{e:mtbar-y}, one can obtain the rapidity dependent average energy, average longitudinal momentum and average of the modulus of the three momentum as:
	\begin{align}
	\bar{E}(y)&=\bar{m}_T(y) \cosh(y), \label{e:Ebar-y}\\
	\bar{p}_z(y)&=\bar{m}_T(y) \sinh(y),\label{e:pzbar-y}\\
	\bar{p}(y)&=\sqrt{\bar{E}(y)^2-m^2}, \label{e:pbar-y} \\
	\bar{p}_T(y) & = \sqrt{\bar{m}_T(y)^2-m^2}.\label{e:pTbar-y}
	\end{align}
	Using these results of the saddle-point integration, the Jacobian term in Eq.~\eqref{e:dndeta} can be written as
	\begin{equation}
	\frac{\bar{p}(y)}{\bar{E}(y)} \equiv J(y) \, = \, \sqrt{1-\frac{m^2}{\bar{m}_T(y)^2 \cosh^2(y)}}.\label{e:J-y}
	\end{equation}
	
	
	Using the above equations, the pseudorapidity can be expressed in terms of rapidity as follows:
	\begin{equation}
	\eta_p (y)  =  \tanh^{-1} \left( \frac{\tanh(y)}{ \sqrt{ \displaystyle\strut 1 - \frac{m^2}{\bar{m}_T(y)^2\cosh^2(y)}}} \right) ,\label{e:eta-y}
	\end{equation}
	This relation is valid for a rapidity dependent average transverse mass $\bar{m}_T(y)$ and in general it cannot be inverted.
	In this case, the parametric expression of the pseudorapidity distribution, given in Eq.~\eqref{e:dndeta} can be used for data fitting.
	
	We have seen above that the average transverse mass $\bar{m}_T(y)$ is, to a leading order, independent of the rapidity $y$,
	if $\lambda - 1 \ll 1$. In this approximation, $\cosh^{\alpha(\kappa,\lambda)}\left(\frac{y}{\alpha(1,\lambda)}\right)\approx 1$, and 
	the average transverse mass becomes rapidity independent, $\bar{m}_T(y) \approx \bar{m}_T(y= 0)$.
	If $\lambda - 1 \ll 1$, we can utilize this approximation to obtain an expicit 
	formula for the pseudorapidity density distribution from the CKCJ solution. 
	
	If $\lambda - 1 \ll 1$  and $\bar{m}_T(y) \approx \bar{m}_T(y= 0)$, the following relations are also valid:
	\begin{eqnarray}
	\bar{p}_T & \approx & \sqrt{(m + T_{\rm eff})^2 - m^2} = M, \label{e:constant-mean-pT} \\
	\bar{p}(\eta_p) & \approx & M \cosh(\eta_p), \\
	\bar{p}_z(\eta_p) & \approx & M \sinh(\eta_p), \\
	\bar{E}(\eta_p) & \approx & \sqrt{m^2 + M^2 \cosh^2(\eta_p)}.
	\end{eqnarray}
	At this point, in order to simplify the subsequent formulas, let us introduce a new variable $D= m/M$,
	with an important meaning to be explained later, when the physical meaning of $D$ becomes clear:
	\begin{eqnarray}
	D^2 &=& \frac{m^2}{M^2}, \\
	J(\eta_p) & = & \frac{\bar{p}(\eta_p)}{\bar{E}(\eta_p)} 
	\, \approx \, 
	\frac{ \cosh(\eta_p) }{\displaystyle \strut \sqrt{ D^2+ \cosh^2(\eta_p) } }, \\
	y(\eta_p) & \approx & \tanh^{-1} \left( \frac{ \cosh(\eta_p) }{\displaystyle \strut \sqrt{ D^2+ \cosh^2(\eta_p) } } \tanh(\eta_p)
	\right) ,\label{e:y-eta}
	\end{eqnarray}
	where $\tanh^{-1}(z)$ is the inverse function of the tangent hyperbolic function.
	
	Using the above equations, we find the explicit formula for the pseudorapidity distribution as follows:
	\begin{equation}
	\frac{dN}{d\eta_p} \approx 
	\frac{\langle N\rangle}{\big(\displaystyle\strut 2\pi\Delta^2 y\big)^{1/2}}
	\frac{\cosh(\eta_p)}{\big(\displaystyle\strut   D^2 + \cosh^2(\eta_p) \big)^{1/2}}
	\exp\Bigg(-\frac{y^2}{ 2 \Delta^2 y \phantom{\Big|} }  \Bigg) \,
	\Bigg|_{y = y(\eta_p)} \label{e:dndeta-function}
	\end{equation}
	where the pseudorapidity dependence of the rapidity is given by Eq.~\eqref{e:y-eta}.
	
	This formula clarifies the physical meaning of the dimensionless parameter $D$. Actually,
	$D^2$ is a dimensionless depression parameter, or dip parameter: it controls the {\it d}epth of the {\it d}ip around midrapidity
	in the pseudorapidity distribution. At large pseudorapidities, the $dN/d\eta_p$ distribution follows the shape
	of the rapidity distribution, but at mid-rapidity, it is {\it d}epressed, 
	up to a factor of ${1/\sqrt{1+D^2}}$, as compared to the rapidity distribution. 
	The rapidity distribution is given as a function in Eq.~\eqref{e:dndy-function}, that is obtained using
	a saddle-point integration technique. Although this advanced calculation suggests that the rapidity density is non-Gaussian,
	near mid-rapidity, this non-Gaussian function can be approximated by a Gaussian shape, as given in  Eq.~\eqref{e:dndy-Gauss}.
	However, to measure the rapidity
	distribution, particle identification is necessary, so in practice the LHC experiments 
	typically publish first the pseudo-rapidity distribution. 
	
	We have evaluated the pseudo-rapidity distribution of the CKCJ solution, too, and our best analytic result, based on a saddle-point
	integration technique  is given as a parametric curve in Eq.~\eqref{e:dndeta}, derived first in Refs.~\citen{Csorgo:2018pxh}. The applications of these formulae were detailed recently in Ref.~\citen{Csorgo:2018fbz}.
	In practice, it  also possible to test, if the Gaussian approximation (to the rapidity distribution, $dN/dy$ ) is satisfactory or not, by
	fitting (the pseudorapidity distribution, $dN/d\eta_p$) using Eqs.~\eqref{e:dndy-Gauss}, \eqref{e:dndy-Gauss-norm}, \eqref{e:y-eta} and \eqref{e:dndeta-function}.
	If such a fit turned out to be not satisfactory for a given dataset and a given experimental precision,
	one can attempt an improved fit using the more precise Eq.~(\ref{e:dndy-function}) together with Eqs.~\eqref{e:y-eta} and \eqref{e:dndeta-function}
	to describe the experimental data on the pseudorapidity distribution.
	Given that  the Gaussian approximation is suspected to fail at LHC energies,  we recommend to start directly with this method, approximating the experimental  rapidity distribution with Eq.~\eqref{e:dndy-function}.

	Phenomenologically, one may fit the pseudorapidity distribution by the explicit but approximate function of Eq.~\eqref{e:dndeta-function} as well.
	On this crudest level
	of approximation, the pseudorapidity distribution depends only on 3 parameters: the normalization $\frac{dN}{dy}\Big|_{y=0}\Big.$,
	the width parameter $\Delta^2 y$, and the dimensionless dip parameter $D^2$ . 
	Similarly, one may test if the rapidity distribution can be approximated, 
	within the experimental uncertainties,  with the Gaussian form of Eq.~\eqref{e:dndy-Gauss}. If such a Gaussian approximation
	describes a certain set of  experimental data, the normalization constant  $\frac{dN}{dy}\Big|_{y=0}\Big.$
	can be expressed in terms of the mean multiplicity $\langle N\rangle$, as given in Eq.~\eqref{e:dndy-Gauss-norm}.
	
	This result, Eq.~\eqref{e:dndeta-function} is yet another  beautiful example of a hydrodynamical scaling behaviour: 
	although the CKCJ exact solution at the time of the freeze-out predicts that the observables will depend on four different parameters,
	$\left.\frac{dN}{dy}\right|_{y=0} $ $\kappa$, $\lambda$ and $T_{\rm eff}$, actually in the $\lambda \rightarrow 1$ limiting case a 
	scaling behaviour is found and the pseudo-rapidity density becomes a function of only three combinations
	of the four physical fit parameters, namely $\langle N\rangle$, $\Delta y$ and $D$.
	
	We expect that these nearly Gaussian  approximations are valid only in a certain limited range of rapidities or pseudo-rapidities, 
	typically up to 2, 2.5 units away from mid-rapidity,  
	as detailed in Refs.~\citen{Csorgo:2018pxh,Csorgo:2018fbz}. We also expect that the 
	range of validity of the more precise Eq.~\eqref{e:dndy-function} is a broader interval in $y$ or in $\eta$, as compared
	to the simplified Gaussian form of Eq.~\eqref{e:dndy-Gauss}. 
	References~\citen{Csorgo:2018pxh,Csorgo:2018fbz} detail the {\it maximum} of the (pseudo)rapidity range,
	where the fits with Eqs.~\eqref{e:dndy-function} and \eqref{e:dndeta} are expected to work. These limits of the domain of the 
	validity of these fits are expressed there in terms of the fit parameters. One of the open questions seems to be that apparently
	the fits seem to work better and in a larger pseudo-rapidity range, as they are expected, as demonstrated 
	on Xe+Xe data at LHC energies in section~\ref{s:4}. 
	
	We may conjecture, that the derivation of these analytic formulae to fit the (pseudo)rapidity
	densities can possibly be generalized and extended to a broader class of exact solutions of relativistic hydrodynamics
	and to a larger range of (pseudo)rapidities. To check this conjecture, it is necessary to search for suitable generalizations or extensions 
	of the CKCJ family of solutions~\cite{Csorgo:2018pxh}.
	However, the detailed investigation of such possible generalizations of an already published exact solution
	goes also well beyond the limitations of
	our current manuscript.
	
	In short, the pseudorapidity distribution of Eq.~\eqref{e:dndeta-function} is fully specified and given as a function
	of $\eta_p$, and the fit range is also well defined and limited to a finite interval around mid-rapidity.
	This Eq.~\eqref{e:dndeta-function} is now given as an explicit function of the pseudo-rapidity.
	Let us emphasize, that this function is obtained from the parameteric curve of
	Eq.~\eqref{e:dndeta} in the  $\lambda - 1 \ll 1$ approximation, that results in the $\cosh^{\alpha(\kappa,\lambda)}\left(\frac{y}{\alpha(1,\lambda)}\right) \approx 1 + {\cal O}((\lambda -1)^2)$ approximation.
	This approximation is very useful, but its validity has to be checked
	when a given data-set is analyzed - in principle, the parametric curve of Eq.~\eqref{e:dndeta} it also implies that the average transverse momentum $\bar{p}_T$
	is approximately independent of the longitudinal momentum component, if Eq.~\eqref{e:dndeta-function} is a valid approximation.
	These formulae are well suited for data fitting, as shown for example on Fig.~\ref{fig:dndeta_xexe}.

	To describe measurements by fits of formulae, let us clarify that Eq.~\eqref{e:dndeta} depends on  four fit parameters that are listed as follows:
	\begin{enumerate}
		\item the average speed of sound $c_s=1/\sqrt{\kappa}$, that corresponds to the equation of state,
		\item the effective temperature $T_{\rm eff}$ that corresponds to the slope of the  $m_T -m$ spectra at mid-rapidity,
		\item the parameter $\lambda$, that controls the relativistic acceleration of the fluid,
		\item and $dN/dy|_{y=0}$, the normalization at midrapidity, that corresponds to the particle density at midrapidity.
	\end{enumerate}
	When comparing to measured $dN/d\eta_p$ data, we take $\kappa$  from the measurement of the PHENIX Collaboration, Ref.~\citen{Adare:2006ti}
	, $T_{\rm eff}$ is also taken from the experimentally determined transverse momentum spectra, 
	so from  fits to the rapidity or pseudorapidity distributions of Ref.~\citen{Alver:2010ck}, one obtains the shape and the normalization parameters $\lambda$ and $dN/dy|_{y=0}$, respectively. See Refs.~\citen{Csorgo:2018pxh, Csorgo:2018fbz} for further details and applications. 
	
	In the Gaussian approximation, the speed of sound or equation of state and the effective temperature
	$T_{\rm eff}$ enter to the fit only through their combination that determines the width parameter $\Delta^2 y$, 
	so instead of two independent fit parameters of $T_{\rm eff}$ and $\kappa$ only their combination $\Delta^2 y$  can be determined
	from the data. 
	Assuming a given value of $\kappa$, the effective temperature $T_{\rm eff}$ can be determined from the
	width parameter $\Delta^2 y$ using Eq.~\eqref{e:dndy-Gauss-width}.
	In this Gaussian approximation, the pseudo-rapidity density can thus be described by three parameters, itemized as follows:
	\begin{enumerate}
		\item the mean multiplicity $\langle N\rangle$ that controls the normalization, 
		\item the width parameter $\Delta y$ that controls the width,
		\item and dip parameter $D$, that controls the  depth of the $dN/d\eta_p$ distribution at mid-rapidity.
	\end{enumerate}
	
	As a closing remark for this subsection, let us emphasize that in the midrapidity region (where $y\approx 0$) 
	one can go beyond the rapidity independent mean $p_T$ approximation of Eq.~\eqref{e:constant-mean-pT}
	by evaluating the leading, second  order correction terms as a function of $y$. 
	The resulting
	rapidity dependent average transverse mass is obtained  to have a Lorentz distribution as follows:
	\begin{eqnarray}
	T_{\rm eff}(y)  & \equiv & \bar{m}_T(y) - m \, = \, \frac{T_{\rm eff}}{1+ a \, y^2}, \label{e:Lorentzian-mt}
	\end{eqnarray}
	where the physical meaning of the effective temperature $T_{\rm eff}$ becomes clear as the mean (thermally averaged) transverse mass at mid-rapidity. The parameter $a$ that controls the width of this Lorentz distribution is found to depend on 
	a combination of the equation of state parameter $\kappa$ and on the acceleration parameter $\lambda$ as
	\begin{eqnarray}
	a & = & 
	\frac{1}{2} \frac{\alpha(\kappa,\lambda)}{\alpha^2(1, \lambda) } .\label{e:a}
	\end{eqnarray}
	Note, that such a   Lorentzian rapidity dependence for the slope of the transverse momentum spectra has been
	derived from analytic hydrodynamics first in Ref.~\citen{Csorgo:1995bi}, assuming a scaling Bjorken flow in the longitudinal
	direction but with a Gaussian density profile. As far as we know, the EHS/NA22 collaboration  has been the first to perform a detailed
	and combined hydrodynamical analysis of single-particle spectra and two-particle Bose–Einstein correlations in high energy physics in
	Ref.~\citen{Agababyan:1997wd}. Although the derivation is valid only
	for symmetric collisions,   a Lorentzian shape of the rapidity dependent average transverse momentum
	has been observed, within errors, by the EHS/NA22 experiment in Ref.~\citen{Agababyan:1997wd} in the slightly asymmetric hadron-proton collisions, using a  mixed $\pi^+ / K^+$  beam with a laboratory momentum of  $250$ GeV/c.
	
	The above result, Eq.~\eqref{e:Lorentzian-mt} is  our third  beautiful example of  hydrodynamical scaling behaviours: 
	although the CKCJ exact solution at the time of the freeze-out predicts that the observables will depend on four different parameters,
	$\left.\frac{dN}{dy}\right|_{y=0} $,
	$\kappa$, $\lambda$ and $T_{\rm eff}$, actually in the $\lambda \rightarrow 1$ limiting case a data 
	collapsing behaviour is found and the rapidity dependent mean transverse momentum
	becomes a function of only on $T_{\rm eff}$ and on  combination of the two different  physical fit parameters $\kappa$ and $\lambda$,
	denoted as $a$ and defined in Eq.~\eqref{e:a}.

	\subsection{Longitudinal HBT-radii}
	\label{ss:Rlong}
	
	As discussed in Ref.~\citen{Csorgo:1995bi}, for a 1+1 dimensional relativistic source, in a Gaussian approximation the relative momentum dependent part of the two-particle Bose-Einstein correlation function is characterized by  (generally mean pair momentum dependent) longitudinal Hanbury-Brown Twiss (HBT) radii. The general expression  in the longitudinally co-moving system (LCMS), where
	the mean momentum of the pair has zero longitudinal component, reads~\cite{Csorgo:1995bi} as
	\begin{equation}
	R^2_{\rm long}=\cosh^2(\eta_x^s)\tau_f^2\Delta \eta_x^2 + \sinh^2(\eta_x^s) \Delta \tau^2.
	\end{equation} 
	Here $\left(\tau_f,\eta_x^s\right)$ stands for the Rindler coordinates of the main emission point of the source in the $(t,r_z)$ plane, which are derived from a saddle-point calculation of the rapidity density, and including $\tau_f$, the lifetime parameter of the fireball. The $\Delta \tau$ and $\Delta \eta_x$ characteristic sizes define the main emission region around the saddle-point at $\left(\tau_f,\eta_x^s\right)$. For $R_{\rm long}$ measurements at midrapidity, where $\eta_x^s \approx 0$, this formula  can be simplified as
	\begin{equation}
	R_{\rm long}=\tau_f \Delta \eta_x .
	\end{equation}
	If the emission function of particles with $y=0$ can be well approximated by a Gaussian shape, with $\Delta \eta_x$ being the width of the space-time rapidity distribution of these particles with vanishing momentum-space rapidity, the longitudinal HBT-radii at $y=0$ can be derived from the CKCJ solutions as we have already shown in Ref.~\citen{Csorgo:2018crb}:
	\begin{equation}
	R_{\rm long}=\tau_f \Delta \eta_x \approx \frac{\tau_f}{\sqrt{\lambda\left(2\lambda-1\right)}} \sqrt{\frac{T_f}{m_T}},
	\label{e:Rlong-ckcj}
	\end{equation}
	where the transverse mass of the particles is denoted by $m_T = \sqrt{m^2 + p_T^2}$, and $T_f$ stands for the freeze-out temperature that can be extracted from the analysis of the transverse momentum spectra. Note that in general
	$T_f < T_{\rm eff}$ as the effective temperature of the transverse momentum spectra contains not only the freeze-out temperature but also radial flow and vorticity effects, see Refs.~\citen{Nagy:2016uiz,Csorgo:2013ksa,Nagy:2009eq}.

	Importantly, Eq.~\eqref{e:Rlong-ckcj} is found to depend on the acceleration parameter $\lambda$ that characterizes the CKCJ solutions, 
	and to be independent of the $\kappa$ parameter, that characterizes the speed of sound and the Equation of State. It is interesting to note that this result corrects the $R_{\rm long}\approx \frac{\tau_f}{\lambda}\sqrt{\frac{T_{f}}{m_T}}$ approximation of Cs\"{o}rg\H{o}, Nagy and Csan\'ad (CNC) in Refs.~\citen{Csorgo:2006ax,Nagy:2007xn}, obtained in the $\kappa\rightarrow 1$ limit. The effect of the accelerating trajectories is corrected by Eq.~\eqref{e:Rlong-ckcj} by a $\lambda\rightarrow \sqrt{\lambda(2\lambda-1)}$ transformation in the CNC estimate for $R_{\rm long}$, Refs.~\citen{Csorgo:2006ax,Nagy:2007xn}.
	In the boost invariant limit ($\lambda\rightarrow 1)$, the longitudinal radii of the CKCJ solution and the CNC solution both reproduce the same Makhlin - Sinyukov  formula of Ref.~\citen{Makhlin:1987gm}. 
	
	\subsection{\label{ss:scaling-of-Rlong} Scaling properties of $R_{\rm long}$ with (pseudo)rapidity density}
	
	Recently, the ALICE collaboration reported~\cite{Adam:2015pya}, that
	the cubed HBT radii  scale with the  pseudorapidity density in a broad energy 
	region including SPS, RHIC and LHC energies and in a broad geometry region including $p+p$, $p+A$
	and $A+A$ collisions.
	Similar scaling laws have been observed by the PHENIX and STAR Collaborations,
	that suggested that each of the main the HBT radii ($i$ = side, out and long) follow the  $R_{i} \propto N_{\rm part}^{1/3}$ 
	scaling in $\sqrt{s_{NN}} = 200$ GeV Au+Au collisions ~\cite{Adler:2004rq,Adams:2004yc}. Earlier, the CERES collaboration
	reported that the HBT volume $R_{\rm side}^2 R_{\rm long}$ is proportional to $N_{\rm part}$, the number of participants~\cite{Adamova:2002wi}.
	Ref. ~\citen{Lisa:2005dd} observed that the $N_{\rm part}$ scaling of the HBT radii can likely be due to the linear correlation
	between $N_{\rm part}$ and the $dN/d\eta_p$ at mid-rapidity. Recently such a proportionality of the HBT radii and the pseudo-rapidity density was demonstrated on a broad class of data from high multiplicity $pp$ through $pA$ and $AA$ collisions
	by the ALICE collaboration in Ref.~\citen{Adam:2015pya}. However, as far as we know, the $R_{\rm HBT}$ $\propto$ 
	$\big(\frac{dN}{d\eta_p}\big)^{1/3}$ scaling law has not been derived so far from any exact solution of (relativistic) fireball hydrodynamics. 
	
	It is thus  a valid question if our solution  can reproduce the energy-independent scaling that corresponds to Fig.~9 of Ref.~\citen{Adam:2015pya} in a straight-forward manner, or not. Let us investigate this question briefly as follows:
	
	From Eq.~\eqref{e:average_N} one finds that the average multiplicity $\langle N \rangle$ depends on the system size $V_x$ and on the freeze-out temperature $T_f$ through the phase space volume. According to Eq.~\eqref{e:Rlong-ckcj}, the freeze-out temperature can be expressed by the longitudinal radius $R_{\rm long}$. As the freeze-out temperature $T_f$ is an $m_T$ independent constant and it can be expressed as a product of two, transverse mass dependent factors,
	we can evaluate this relation also at the average transverse mass $\overline{m}_T$ to find
	\begin{equation}
	T_f = \frac{m_T \lambda\left(2\lambda-1\right)}{\tau_f^2}R_{\rm long}^2 \, = \, \frac{\bar{m}_T \lambda\left(2\lambda-1\right)}{\tau_f^2}\bar{R}_{\rm long}^2 .
	\end{equation}

	Therefore from Eqs.~\eqref{e:average_N} and \eqref{e:Vp_limit} it is easy to see that the mean multiplicity $\langle N \rangle$ depends on the ($m_T$ averaged) longitudinal radius, $\bar{R}_{\rm long}$ as:
	\begin{equation}
	\langle N \rangle \propto V_x T_f^{3/2} \propto V_x \bar{R}_{\rm long}^3.\label{e:N-Rlong}
	\end{equation}
	Note, that the mean multiplicity $\langle N\rangle$ also depends on the system size, as it is proportional to the total geometrical volume of the fireball at the mean freeze-out time. Note also that the constant of proportionality $T_f^{3/2}$ depends on the  {\it local} freeze-out temperature $T_f$, as it is the measure of the size of the momentum space in the local comoving system.
	
	The proportionality of the pseudorapidity density and $R_{\rm long}^3$, as discussed in Ref.~\citen{Adam:2015pya}, is straigthforward to derive from the CKCJ solution. This proportionality directly follows from the proportionality of Eq.~(\ref{e:N-Rlong}) and
	the relation 
	\begin{equation}
	\frac{dN}{d\eta_p}=\langle N \rangle \rho(\eta_p),
	\end{equation}
	where $\rho\left(\eta_p\right)$ is the normalized pseudorapidity density, with the following normalization condition:
	\begin{equation}
	\int\limits_{-\infty}^{\infty} \rho(\eta_p)d\eta_p=1.
	\end{equation}
	Thus the pseudorapidity distribution is proportional to the mean multiplicity (times a probability density).
	As the mean multiplicity depends on the (transverse mass averaged) longitudinal HBT-radii as well as on the average transverse mass,
	we find that
	\begin{equation}
	\frac{dN}{d\eta_p}\Big|_{\eta_p = 0} =\langle N \rangle \rho(\eta_p = 0) \propto  \bar{R}_{\rm long}^3 .
	\end{equation}
	In other words, at mid-rapidity, $\eta_p \approx 0$, 
	\begin{equation}
	\bar{R}_{\rm long} = A  \Big(\frac{dN}{d\eta_p} \Big)^{\frac{1}{3}}, \label{e:Rscaling}
	\end{equation}
	where the constant of proportionality $A$ is actually a complicated function of
	quantities that may depend on the expansion dynamics in a non-trivial manner. It is clear that $A$ depends
	on $V_x/\tau_f^3$, on the acceleration parameter $\lambda$, on the widht of the rapidity distribution $\Delta y$
	hence on the (temperature averaged) speed of sound $c_s^2 = 1/\kappa$, on the effective temperature $T_{\rm eff}$, on the mass of the particles $m$ and on the average phase-space occupancy $\bar f$. A good strategy can be  to take the value of $A$ directly from measurements.
	Experimentally,  $A$ is found to be weakly dependent on the system size, increasing for larger systems. We need to do more detailed data fitting to cross-check the properties of $A$, than doable within the limits of the current investigations.
	
	For the present paper, let us emphasize only that the proportionality of Eq.~\eqref{e:Rscaling} is a natural consequence of the investigated CKCJ solution of Ref.~\citen{Csorgo:2018pxh}. This is due to the fact that the local freeze-out temperature determines
	the size of homogeneities at the time of freeze-out, when the geometrical sizes of the expanding system is large
	(Refs.~\citen{Makhlin:1987gm,Csorgo:1995bi}), $R_{HBT}\propto T_f^{1/2}$. The same local freeze-out temperature on the power of 3/2, $T_f^{3/2}$ measures the local volume of the momentum space which multiplied together with the total volume of the fireball is proportional to the
	mean multiplicity, hence the pseudorapidity density. Thus the transverse mass averaged HBT radii scale as 
	$\bar{R}_{HBT} = A \Big(\frac{dN}{d\eta_p}\Big)^{\frac{1}{3}}$, for expansion dominated systems with larger geometrical than thermal
	length-scales, as noted in Ref.~\citen{Csorgo:1995bi}. This scaling is expected to break down at low energies where the expansion effects are less important and the geometrical sizes start to dominate the HBT radii.
	
	The above result is formulated in a general manner, and it is valid in a broad class of 1+1 and 1+3 dimensional exact solutions of (relativistic) hydrodynamics. For example, it is valid for several classes of 1+ 1 dimensional relativistic hydrodynamical solutions, including the CKCJ solution of Refs.~\citen{Csorgo:2018pxh,Csorgo:2018fbz}
	and the CNC solution of Refs.~\citen{Csorgo:2006ax,Nagy:2007xn}.
	Although the Bjorken-Hwa solution of Refs.~\citen{Bjorken:1982qr,Hwa:1974gn} leads to a flat rapidity density distribution,
	it can be considered as the $\lambda \rightarrow 1$ limiting case of our description. Hence the scaling law of the
	longitudinal HBT radius as expressed by  Eq.~\eqref{e:Rscaling} is proven here to be valid for the boost-invariant 
	Hwa-Bjorken solution, too.
	
	We also find,  that the validity of this derivation can be extended 
	to the non-relativistic kinematic domain, where $m_T \approx m$,
	and the same result is obtained for the 1+ 3 dimensional, non-relativistic exact solutions of fireball hydrodynamics of Refs.~\citen{Csizmadia:1998ef,Csorgo:2001xm,Csorgo:1998yk,Csorgo:2013ksa,Csorgo:2015scx}, extending the 
	derivation not only to the longitudinal HBT radius but also to the transverse (side, and out) HBT radii.
	
	Our derivation provides  a novel method that can be applied in a broader and more general
	class of exact solutions of fireball hydrodynamics than the considered 1+1 dimensional CKCJ solution.
	It seems to us that the key steps are the proportionality of the mid-rapidity pseudorapidity density to the average HBT volume
	(evaluated at mid-rapidity and at the average value of the transverse momentum)
	times the  volume of the invariant  momentum distribution (in momentum-space). 
	Another way to consider this result is to note that when evaluating the 
	mean multiplicity, the order of the invariant integration over the momentum distribution and the integration
	over the freeze-out hypersurface is exchangeable.
	

	A  detailed exploration of this topic and the gerenalization of our method to other exact solutions of relativistic hydrodynamics  looks indeed very promising and attractive. A more detailed proof of these deep properties of exact solutions of fireball hydrodynamics goes, however, well beyond the scope of our present manuscript, that
	focusses on the exploration of the experimental consequences and observable relations that can be derived from the 
	CKCJ family of exact solutions of relativistic fireball hydrodynamics in 1+1 dimensions.  
	
	\section{Data analysis and fit results}\label{s:4}
	In what follows, let us detail and explain, how one can improve on  Bjorken's estimate for the initial energy density, using straightforward fits to the already published data and relying on exact results from 1+1 dimensional relativistic hydrodynamics.
	
	To extract $\tau_f$, the lifetime parameter of the fireball, we fit the pseudorapidity density to extract the shape parameter $\lambda$, and the longitudinal HBT-radius of the CKCJ solution to the experimental data. For  comparisions of the consequences of the CKCJ hydro solutions  with measurements, we have selected data on $\sqrt{s_{NN}} = $ 130 GeV and 200 GeV  Au+Au  collisions, because  the pseudorapidity density, the slope parameter of the transverse momentum spectra and the HBT radii are known to be measured in the same, 0-30\% centrality class. In both of these reactions, the PHENIX collaboration published their two-pion correlation results in the 0-30\% centrality class, consequently we fitted our hydrodynamic model to the pseudorapidity density data of PHOBOS experiment from Ref.~\citen{Alver:2010ck}, that we averaged in this 0-30 \% centrality class. The centrality dependence of the effective temperature can be estimated with the help of an empirical relation,
	\begin{equation}
	T_{\rm eff}=T_f + m\langle u_T \rangle^2.  \label{e:teff}
	\end{equation}
	This  relation has been derived  for a three-dimensionally expanding, ellipsoidally symmetric exact solution of fireball hydrodynamics
	with a temperature dependent speed of sound, as a model for non-central heavy ion collisions~\cite{Csorgo:2001xm}, 
	but only for a single kind of hadron that had a fixed mass $m$.
	Recently this relation has also been  re-derived for a non-relativistic, 3 dimensionally expanding fireball that hadronizes to
	a mixture of various hadrons that have different masses $m_i$ just like a mixture of pions, kaons and protons, as detailed in Ref.~\citen{Csorgo:2018tsu}. This derivation included a lattice QCD EoS in the quark matter or strongly interacting quark gluon plasma phase~\cite{Csorgo:2018tsu}. 
	
	It is remarkable that the rotation or vorticity of the nearly perfect fluid of Quark-Gluon Plasma~\cite{STAR:2017ckg} does not change the affine-linear relation between the effective temperature and the mass of particles. As it was shown in Ref.~\citen{Csorgo:2015scx} the transverse flow $\langle u_T\rangle$ in Eq.~\eqref{e:teff} can be identified with a quadratic sum of the radial flow $\dot{R}_f$
	and the tangential flow that corresponds to the rotation of the fluid, $\omega_f R_f$ as
	\begin{equation}
	\langle u_T \rangle^2 = \dot{R}_f^2 + \omega_f^2 R_f^2  ,  \label{e:ut}
	\end{equation}
	where at the time of the freeze-out, $R_f$ is the transverse radius of an expanding and rotating fireball,
	$\dot{R}_f$ stands for the radial flow ie the time derivative of $R(t)$ at the freeze-out time $t_f$ and $\omega_f$ is
	the angular velocity of the rotating fireball at the freeze-out time $t_f$. This relation was derived, as far as we know, for the first time in Ref.~\citen{Csorgo:2015scx},
	using a lattice QCD motivated equation of state, for a general temperature dependent speed of sound given by $\kappa(T)$,
	assuming a  spheroidal symmetry and non-relativistic kinematics. This derivation was generalized to a triaxial, rotating and expanding fireball in Ref.~\citen{Nagy:2016uiz}, that found that $R_f = 0.5 (X_f + Z_f)$ is the average freeze-out radius for a fireball
	rotating around the transverse ($r_y$) axis and off-diagonal terms correspond to the difference of the rotating and expanding motion of the fluid within the triaxial ellipsoid and the rotating and expanding motion of geometrical shape of the triaxial ellipsoid~\cite{Nagy:2016uiz}.

\begin{figure}[h!] 
		\includegraphics[scale=0.425]{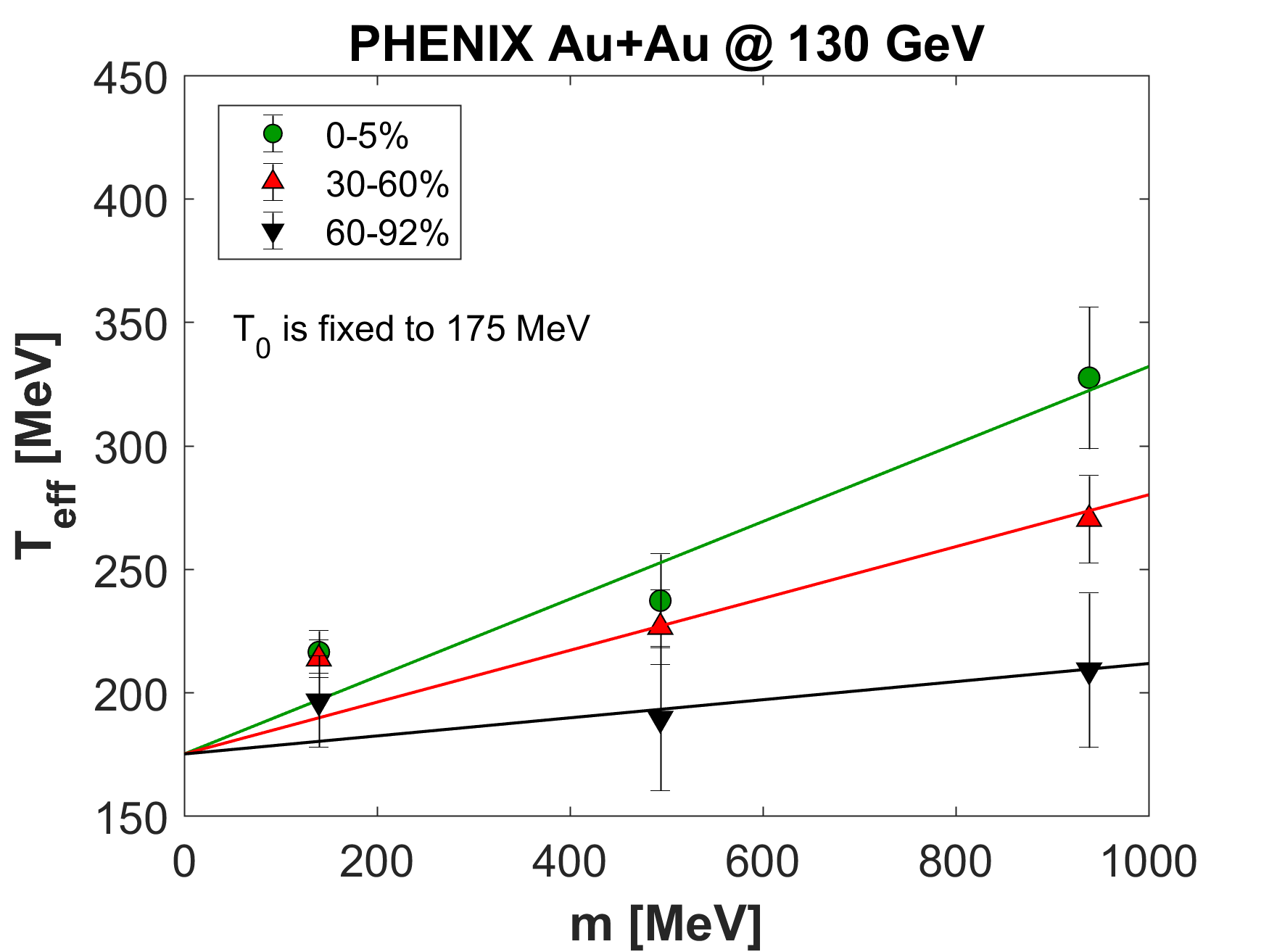} 
		\includegraphics[scale=0.425]{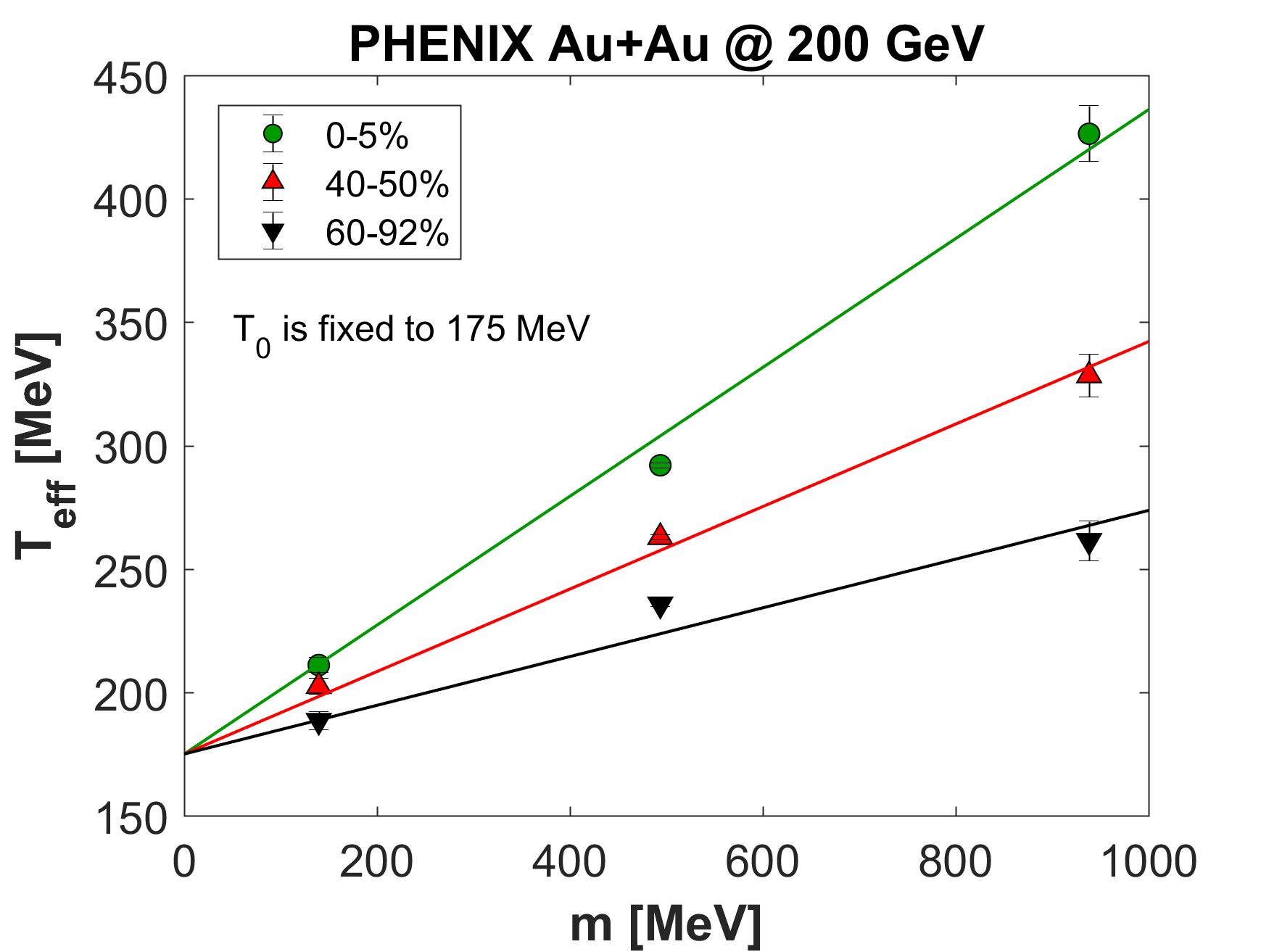} 
		\centering 
		\caption{The mass and centrality dependence of the effective temperature of charged pions and kaons, as well as protons and anti-protons in $\sqrt{s_{NN}}=130$ GeV (Ref.~\citen{Adcox:2003nr}) and $\sqrt{s_{NN}}=200$ GeV (Ref.~\citen{Adler:2003cb}) Au+Au collisions. For each centrality classes, the linear fits are shown by continuous lines. The freeze-out temperature parameter $T_f$ is fixed to 175 MeV in both cases, although at 130 GeV, a slightly higher $T_f\approx 180$ MeV is preferred by the data.}
		\label{fig:inverse_slope} 
	\end{figure}

	Due to both radial flow and vorticitiy effects, $\langle u_T\rangle$, the slope of the affin-linear function in Eq.~\eqref{e:teff}
	is expected (and experimentally found) to be significantly centrality dependent. On the other hand, the intercept
	of this affine-linear function, $T_f$  is interpreted as the freeze-out temperature, that is expected
	(and experimentally found) to be centrality independent. The numerical value of $T_f$ is empirically found from the analysis of the
	transverse momentum spectrum and it is indeed approximately independent of the particle type, the centrality and the center of mass energy of the collision. Figure~\ref{fig:inverse_slope} illustrates the linear fits to the effective temperature of charged pions, charged kaons, and protons, antiprotons for PHENIX Au+Au at $\sqrt{s_{NN}}$=130 GeV and 200 GeV data, and suggests that the $T_f$ value that is consistent with this
	interpretation of the transverse momentum spectra is  $T_f\approx 175 $ MeV for both $\sqrt{s_{NN} } = 130$ and $200$ GeV Au+Au collisions,  independently of particle mass, centrality and center of mass energy of the collision indeed. 
	Important theoretical expectations suggests a lower value for the freeze-out temperature, corresponding to the mass of the lightest strongly interacting quanta with no conserved charge, $T_f \approx m_{\pi} $, the pion mass.  Thus we have also investigated if $T_f = 140$ MeV can be utilized for the 
	extraction of the life-time of the reaction. Lower freeze-out temperatures correspond to larger life-times for the same 
	datasets according to Eq.~\eqref{e:Rlong-ckcj}, and we shall subsequently see  that longer life-times correspond to larger initial energy densities for the same pseudorapidity distributions. In what follows, we utilize the more conservative,  shorter life-times obtained with larger freeze-out temperatures in Table~\ref{t:freezeout}. These values give lower estimates for the initial energy densities. However,  we also test the effects the possible longer lifetimes and larger initial energy densities, corresponding to the lower value of $T_f = 140$ MeV. We  summarize the corresponding longer freeze-out times  also in Table~\ref{t:freezeout}, so that the interested reader may evaluate the corrections that originate from the choice of a lower freeze-out temperature to the initial energy density estimates in a straightforward manner.

	If the effective temperature is allowed to be a free fit parameter in the $dN/d\eta_p$ fits, we find, that the fit results are in agreement with the estimated centrality dependence of $T_{\rm eff}$ that are determined from Fig.~(\ref{fig:inverse_slope}).
	As the pseudorapidity distributions depend not only on the effective temperature $T_{\rm eff}$ but also on the average
	speed of sound, we fixed the speed of sound to $c_s^2=0.1$ in accordance with the experimental results of the PHENIX collaboration~\cite{Adare:2006ti}. Since the CKCJ solution is valid in a limited, central pseudorapidity interval only, we fitted the $\frac{dN}{d\eta_p}$ of the CKCJ solution to data in the $\left[-2.5,2.5\right]$ pseudorapidity interval. The fits and the best fit parameters are shown on Figure~\ref{fig:pseudorapidity-density}. 
	
	It is important for the current study, that $T_{\rm eff}$, $\frac{dN}{d\eta_p}$ and $R_{\rm long}$ must be determined in the same centrality class of the same colliding system at the same energy. As it can be seen on the right panel of Fig.~\ref{fig:inverse_slope}, for Au+Au at $\sqrt{s_{NN}}$=200 GeV, $T_{\rm eff}$ data are available  in the 0-5 \% and 40-50 \% centrality classes, but they were not directly measured in the 0-30 \% centrality class. Note that the $T_{\rm eff}=$203 MeV value that we obtained from fitting the $\frac{dN}{d\eta_p}$ data at $\sqrt{s_{NN}}$= 200 GeV  for the 0-30 \% centrality class is just in between the values that are obtained  for the 0-5 \% and 40-50 \% centrality classes, as indicated on Fig.~\ref{fig:inverse_slope}. However, for the same system at $\sqrt{s_{NN}}$=130 GeV, $T_{\rm eff}$ data are measured also in the 5-15 \% and the 15-30 \% centrality classes -- although these are not shown on the left panel of Fig.~\ref{fig:inverse_slope}. We evaluated the inverse slope for the 0-30 \% centrality class by averaging the fit of the $T_{\rm eff}=T_f+m\langle u_T \rangle^2$ relation over the 0-5 \%, 5-15 \% and 15-30 \% centrality classes for the pion mass, and obtained values that are consistent with both the corresponding values from the $m_T - m $ spectrum fits,
	as indicated on Fig.~\ref{fig:inverse_slope}, and from the fits of the pseudo-rapidity distributions as discussed in the next sub-section.
	
	\subsection{Fits to the pseudorapidity distribution}
	
	We describe below two kind of pseudo-rapidity distributions: first, we detail fits to $Au$+$Au $ collision data at the RHIC energies
	at $\sqrt{s_{NN}} = 130$ and $200$ GeV, then we discuss fits to recent CMS data on $Xe$+$Xe$ collisions at
	$\sqrt{s_{NN}} = 5. 44 $ TeV at LHC.
	
	\subsubsection{\it Fits to PHOBOS data in Au+Au collisions at RHIC}
	The PHOBOS collaboration published an extensive study of pseudorapidity distributions in $Au$+$Au$ collisions at various energies at RHIC~\cite{Alver:2010ck}.
	Unfortunately, the point-to-point fluctuating statistical and systematic errors of the $\frac{dN}{d\eta_p}$ data are not separated from the point-to-point correlated normalization error in these publications~\cite{Alver:2010ck}. This makes a simple $\chi^2$ test practically useless, as it leads to overly small estimations for the value of $\chi^2$. We have analyzed these data
	with an advanced $\chi^2$ test, described as follows.
	
	As a preparatory and necessary control point, we have fitted the PHOBOS pseudorapidity distributions with our formulae 
	of Eq.~\eqref{e:dndeta},  using the uncertainty band, given by the PHOBOS experiment in Ref.~\citen{Alver:2010ck}
	as a preliminary error estimate. Unfortunately this uncertainty band did not separate the point to point fluctuating errors from the point-to-point correlated systematic and overall correlated systematic errors. The fits were excellent from the point of view of the $\chi^2/NDF$ however they were not sufficiently well constrained to follow the shape of the data points, as the overall errors
	were apparently so large that the shape was not constrained too much. Because of these circumstances, we tried to estimate the point-to-point fluctuating errors and a more stringent and  advanced estimated for value of $\chi^2$. 
	Consequently, we have fitted the $\frac{dN}{d\eta_p}$ data in three steps, as follows: 
	
	In the  first step, tried to create a smooth line that interpolates between the datapoints. In order to achieve this, we used an estimated (ad hoc) $\eta_p$ independent, 3 \% fluctuating uncorrelated uncertainty on each datapoint and we used these errors to minimize the value of $\chi^2$. Using  the fitted parameters, obtained from this first step,  we got a good-looking overall description that
	we considered as a smooth interpolating line that smoothly connects the data points, in a reasonable looking manner. 
	
	In a second step, we estimated the sum of the point-to-point fluctuating error of each data points of the $\frac{dN}{d\eta_p}$ spectra by determining the deviations of the data points from the interpolating line determined in the previous, first step, 
	and we  considered these deviations as our best estimations $e_i$ for the fluctuating errors of data points $d_i$. 
	
\begin{figure}[h!] 
		\includegraphics[scale=0.425]{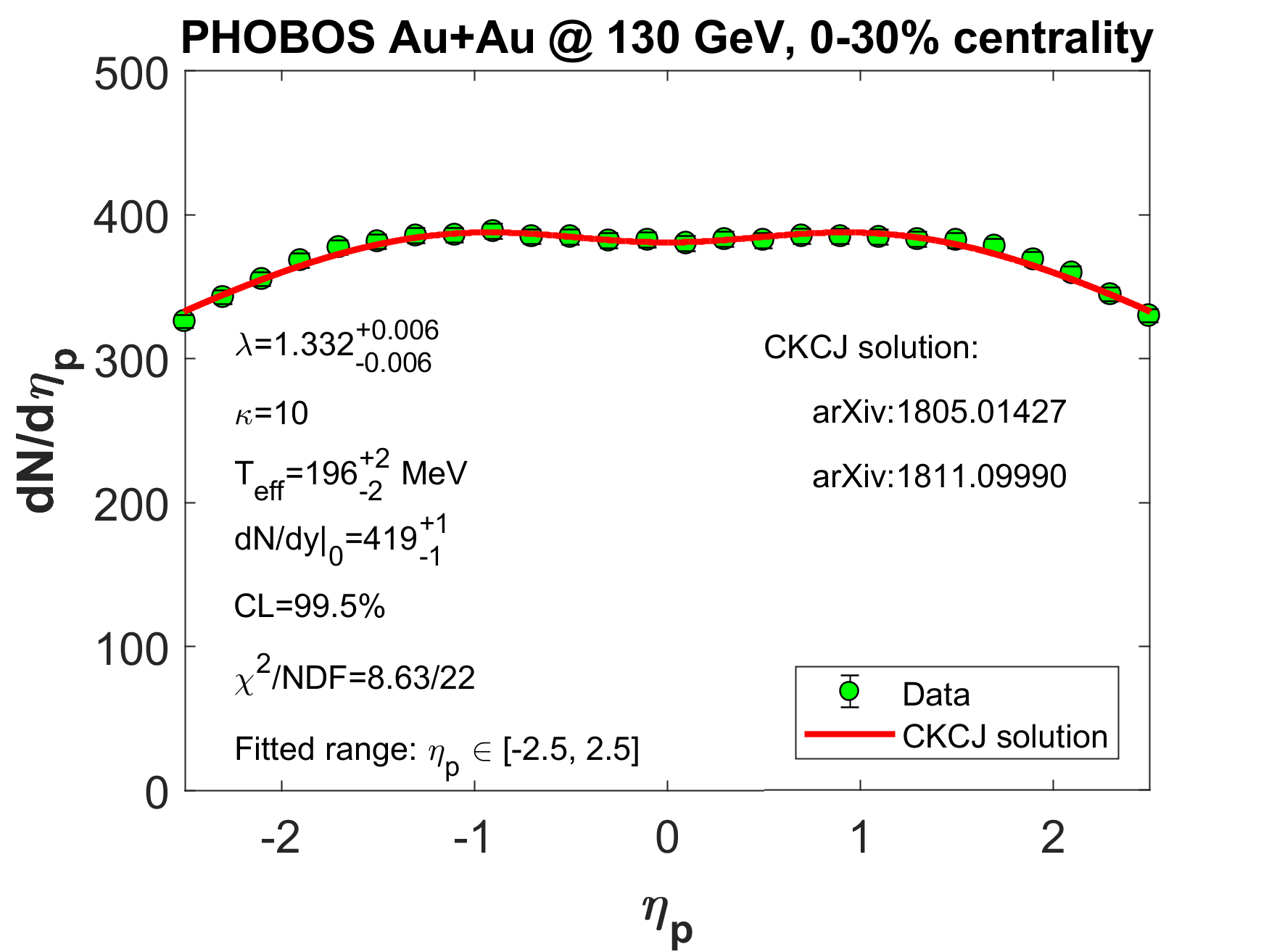} 
		\includegraphics[scale=0.425]{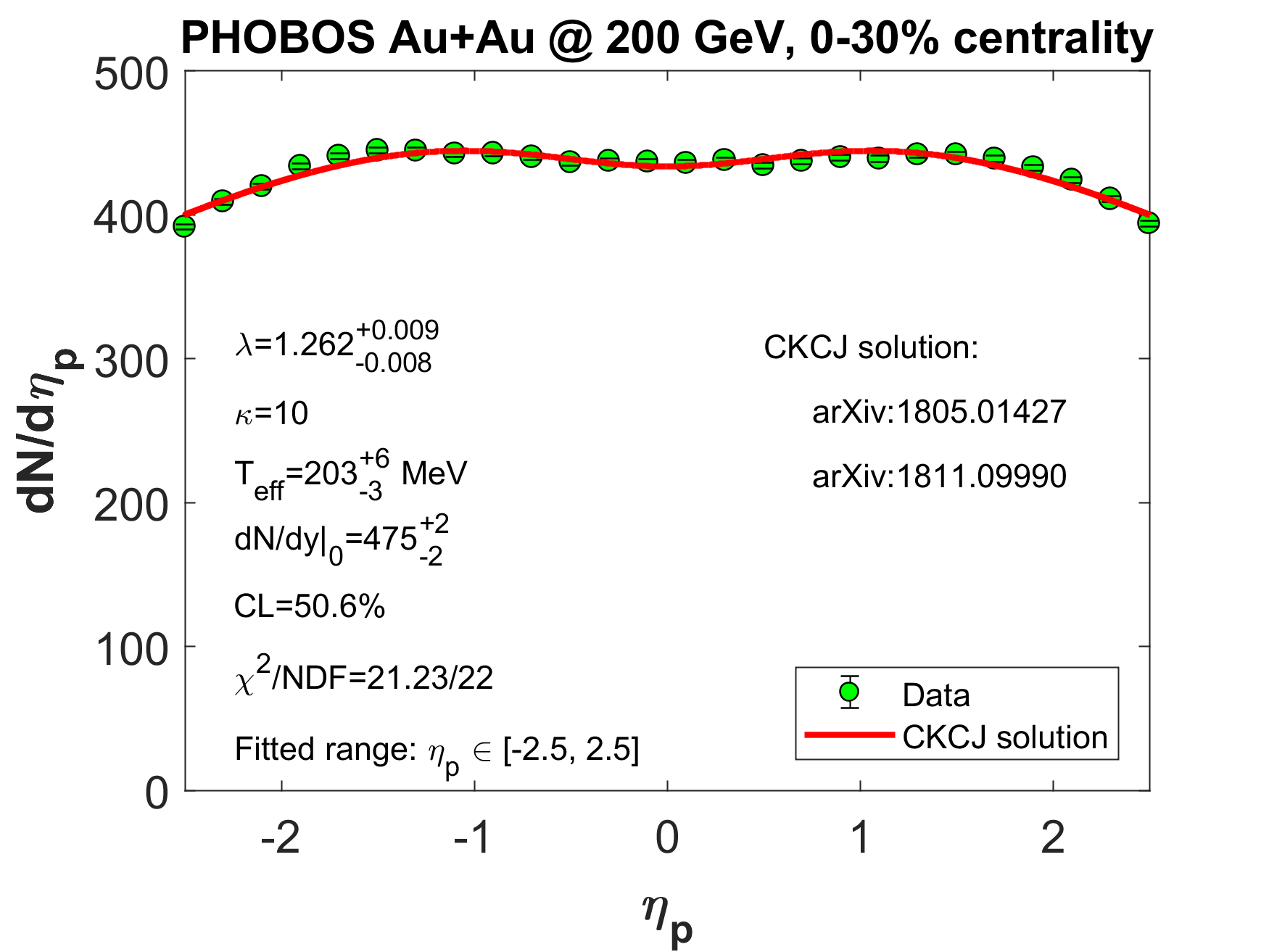} 
		\centering 
		\caption{Fits of the pseudorapidity density with the CKCJ hydro solution~\cite{Csorgo:2018pxh}, to PHOBOS Au+Au data at $\sqrt{s_{NN}} = $ 130 GeV (Ref.~\citen{Alver:2010ck}) (left) and $\sqrt{s_{NN}} = $ 200 GeV (Ref.~\citen{Alver:2010ck}) (right) in the 0-30 \% centrality class. The fit quality is determined by $\chi^2$ defined by Eq.~\eqref{e:chi2} and, as usual, the number of degrees of freedom is denoted by the acronym $NDF$.  The speed of sound is $c_s^2 = 1/\kappa = 0.1$, fixed in both cases. The large systematic errors on PHOBOS data are suppressed on this plot.}
		\label{fig:pseudorapidity-density} 
	\end{figure}

	In the third step,  we have repeated the $\chi^2$ minimization with the data points $d_i$ and our estimated point-to-point fluctuating errors $e_i$. This is the value of $\chi^2$ and the corresponding confidence level CL
	that we report on Figure~\ref{fig:pseudorapidity-density}.  
	Thus the $\chi^2$ values, for the advanced calculations, indicated on Figure~\ref{fig:pseudorapidity-density} are defined as:
	\begin{equation}
	\chi^{2}=\sum\limits_i \frac{\Big(d_i - f\big(\eta_{p,i}\big| \big. \kappa, \lambda, T_{\rm eff},	\big.\frac{dN}{dy}({y=0})\big)\Big)^2}{e_i^2}, \label{e:chi2}
	\end{equation}
	where $f\big(\eta_{p,i}\big| \big.\kappa, \lambda,T_{\rm eff}, 	\big.\frac{dN}{dy}({y=0})\big)$ 
	is the theoretical result given as a parameteric curve, corresponding to 
	Eqs.~\eqref{e:dndy-function}, \eqref{e:dndeta} and \eqref{e:mtbar-y}-\eqref{e:y-eta} 
	with $\kappa = 10$ fixed values and using mid-rapidity density as a normalization.
	Given  that the values of $e_i$ defined above correspond to the actual point-to-point fluctuations of the PHOBOS data points, that  are significantly smaller as compared to the total and point-to-point correlated errors reported by PHOBOS, so our fits are required to follow the data points rather closely. A quality description of the data points can be obtained this way, as indicated on Figure~\ref{fig:pseudorapidity-density}.
	
	As discussed in the previous section, in a Gaussian approximation only a combination of the effective temperature $T_{\rm eff}$
	and the equation of state parameter $\kappa$ enters the fitting function denoted as $f_G$, and another combination of 
	the mass $m$ and the effective temperature, $D$ enters the fitting function. Hence  the
	$\chi^2_G$ distribution, that corresponds to the Gaussian rapidity distributions,  is defined as follows:
	\begin{equation}
	\chi^{2}_G=\sum\limits_i \frac{\Big(d_i - f_G\big(\eta_{p,i}|\langle N\rangle , \Delta y, D\big)\Big)^2}{e_i^2}, \label{e:chi2_2}
	\end{equation}
	where $f_G\big(\eta_{p,i}|\langle N\rangle , \Delta y, D\big)$ 
	is the theoretical curve in a Gaussian rapidity density approximation, 
	corresponding to Eqs.~\eqref{e:dndeta-function}, \eqref{e:dndy-Gauss} and \eqref{e:dndy-Gauss-norm}.
	
	The new results indicated on Figure~\ref{fig:pseudorapidity-density} continue the trend that was demonstrated already in Refs.~\citen{Csorgo:2018pxh,Csorgo:2018fbz}: the lower the collision energy, the greater the acceleration parameter $\lambda$, and the greater the difference from the asymptotic, boost-invariant $\lambda = 1$ limit. Note that relativistic acceleration vanishes in the boost-invariant limit, which corresponds to $\lambda = 1$.
	
	Since we have obtained the effective temperature $T_{\rm eff}$ and the acceleration parameter $\lambda$ from the transverse momentum spectra and the pseudorapidity density data, the number of fit parameters in Eq.~\eqref{e:Rlong-ckcj}, i.e. in our new formula for the longitudinal HBT radii is reduced  to one: the remaining free parameter is the lifetime parameter $\tau_f$. 
	
	\subsubsection{\it Fits to CMS data on Xe+Xe collisions}
	In Ref.~\citen{Sirunyan:2019cgy}, CMS collaboration published  the pseudorapidity density of Xe+Xe collisions measured at $\sqrt{s_{NN}}=5.44$ TeV in the 0-80 \% centrality class. 
	In addition, CMS compared these data to various model calculations, that were to a smaller or greater extent also influenced by numerical solutions of relativistic hydrodynamics. In particular, the comparisons included  the prediction of EPOS LHC of Refs.~\citen{Werner:2005jf,Pierog:2013ria}, that includes a   proper hydro treatment for heavy ion collisions.
	In case of $p+p$ collisions at LHC energies,  a realistic treatment of the hydrodynamical evolution with proper hadronization such an effect was not observed in EPOS, hence the EPOS LHC model includes a non-hydrodynamically generated radial flow in small systems, like $p$+$p$ collisions. 
	The HYDJET Monte Carlo simulation of Ref.~\citen{Lokhtin:2005px} relies on a  superposition of a soft hydro-type state and hard multi-jets.
	The AMPT model includes the Heavy Ion Jet Interaction Generator (HIJING) for generating the initial conditions, Zhang's Parton Cascade (ZPC) for modeling partonic scatterings, the Lund string fragmentation model or a quark coalescence model for hadronization, and A Relativistic Transport (ART) model for treating hadronic rescatterings, in an  improved and combined way to 
	give a coherent description of the dynamics of relativistic heavy ion collision~\cite{Lin:2004en}. 
	Apparently, each of these models had difficulties in predicting the CMS Xe+Xe data at  their present phase of development.
	
	CMS data from Ref.~\citen{Sirunyan:2019cgy}  provided strikingly significant deviations each of these models, although
	each of them included a detailed simulation of resonance decays as well, while we use the core-halo model of Ref.~\citen{Csorgo:1994in} to handle phenomenologically the effect of long-lived resonances. A key property of the core-halo model 
	of Ref.~\citen{Csorgo:1994in} is that provides an understanding and an explanation of why the hydrodynamical scaling properties can be observed in high energy heavy ion and hadron-hadron interactions even if the hydrodynamically evolving core of the reaction is
	surrounded by a halo of (long-lived) resonance decays. Actually the EPOS LHC calculations of Refs.~\citen{Werner:2005jf,Pierog:2013ria} also separate the hydrodynamically evolving core from the halo of resonance decays in a more detailed, but qualitatively similar manner to the core-halo model, but they use the core-corona terminology  ~\cite{Werner:2005jf,Pierog:2013ria} instead of the core-halo terminology of ~\cite{Csorgo:1994in}, to describe apparently similar physics ideas.
	
\begin{figure}[h!] 
		\includegraphics[scale=0.85]{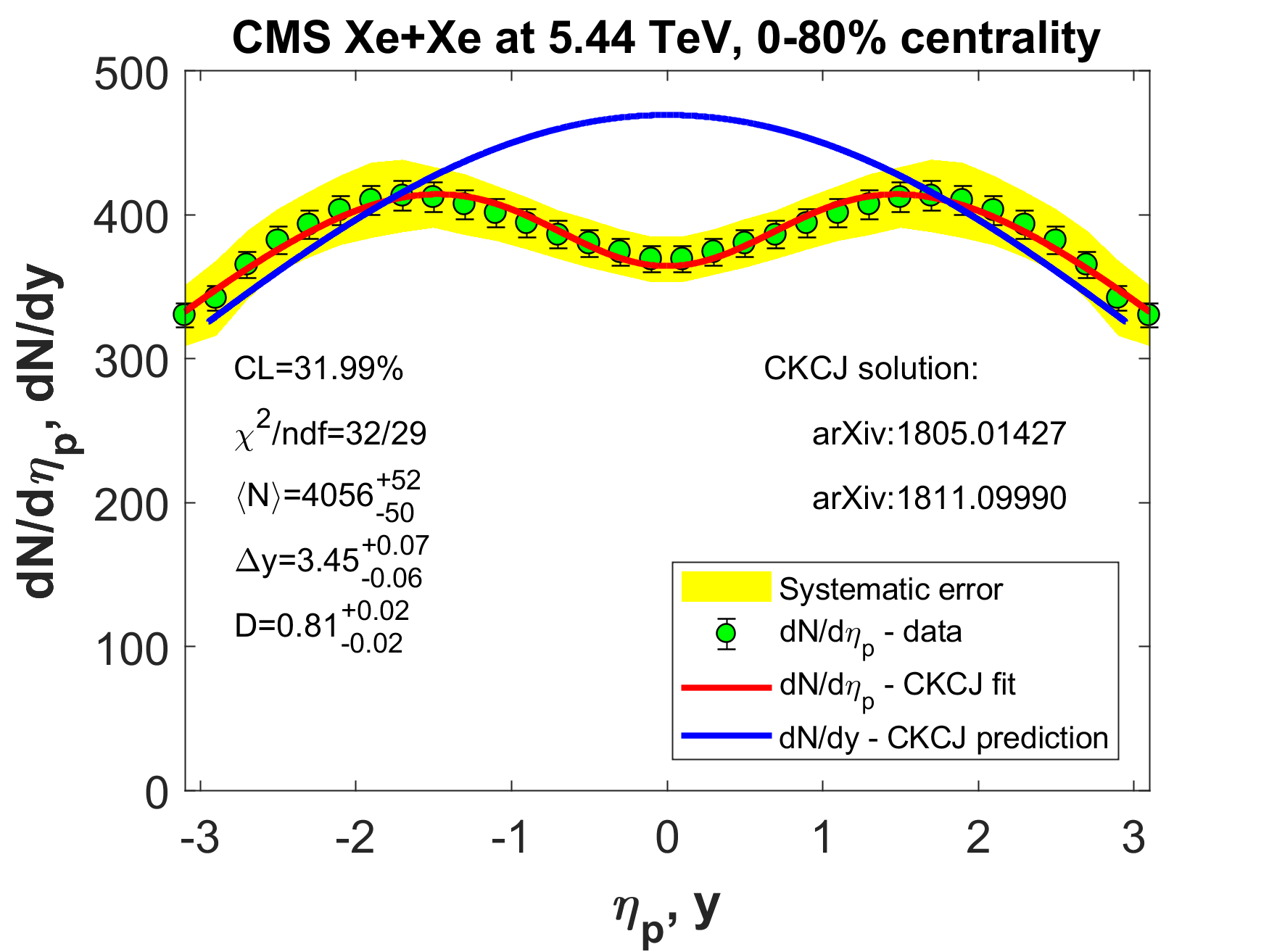}
		\centering 
		\caption{Fits of the pseudorapidity density with the CKCJ hydro solution~\cite{Csorgo:2018pxh}, to CMS Xe+Xe data at $\sqrt{s_{NN}} = $ 5.44 TeV (Ref.~\citen{Sirunyan:2019cgy}) in the 0-80 \% centrality class. 
			This fit is obtained by the strongest, Gaussian approximation for the rapidity density, corresponding to Eqs.~(\ref{e:dndy-Gauss},\ref{e:dndy-Gauss-norm},\ref{e:dndeta-function}). 
			The fit parameters are the mean multiplicity $\langle N\rangle$, the Gaussian width of the pseudorapidity distribution $\Delta y$
			and the dimensionless dip parameter $D$. The fine CKCJ fit quality, the fit range (that includes all the datapoints), and the best value of each of the fitted parameters is indicated on the plot. Yellow band indicates the systematic errors of the CMS measurement. This fit describes the data in a larger pseudorapidity
			domain than expected, as detailed in the body of the manuscript. We have also evaluated the CKCJ prediction for the rapidity density using the same set of parameters, and overlayed the plot of the rapidity density on that  of the pseudorapidity density, corresponding to the same physics but to different measurables.
		}
		\label{fig:dndeta_xexe} 
	\end{figure}

	As indicated on Figs. 1 and 2 of Ref.~\citen{Sirunyan:2019cgy}, each of the AMPT 1.26t5, EPOS LHC v3400, and HYDJET 1.9 models
	fails to describe the pseudorapidity distribution of Xe+Xe collisions at $\sqrt{s} = 5.44$ TeV already on a qualitative level.
	However, we were able to describe  these CMS data even  with the crudest, nearly Gaussian approximation of the CKCJ solution~\cite{Csorgo:2018pxh} of relativistic hydrodynamics, not only on a  qualitative but also on a quantitative level,
	as indicated on Fig.~\ref{fig:dndeta_xexe}. In fact,  the CKCJ  solution~\cite{Csorgo:2018pxh} describes this recent CMS dataset  surprisingly well, as detailed below. 
	
	The fit quality is statistically not unacceptable, actually our three-parameter formula   of Eq.~(\ref{e:dndeta}) provides a fine description of these CMS data.
	We have tested that fitting with the more precise parametric curve of Eq.~(\ref{e:dndeta}), 
	or using the Gaussian approximation of  Eqs.~\eqref{e:y-eta} and \eqref{e:dndeta-function},
	we obtain similarly good fit quality. 
	From the fit parameters of the parameteric curve of Eq.~(\ref{e:dndeta}), using the results of Refs.~\citen{Csorgo:2018pxh,Csorgo:2018fbz}, we have expected that the parameteric curve fit
	is able to describe these Xe Xe data in a rather limited pseudo-rapidity range, corresponding approximately to $-2.2 < \eta_p < 2.2$.
	In fact, Fig.~\ref{fig:dndeta_xexe} indicates that even in the Gaussian approximation,  our fit actually describes the data 
	significantly better, than expected.
	
	In particular, as indicated by Fig.~\ref{fig:dndeta_xexe}, the fit quality is satisfactory for the 
	range of the data description, corresponding to $-3.2 < \eta_p < 3.2$, without comprimising the fit quality.
	This range is significantly extended, as compared to the expected $-2.2 < \eta_p < 2.2$ fit range.
	This is particularly interesting as we have used a hydrodynamical scaling description, where 
	only three combinations ($\langle N\rangle$ , $\Delta y$, $D$) of the four hydrodynamical fit parameters
	$T_{\rm eff}$, $\lambda$, $\kappa$, and $dN/dy|_{y=0}$  are used in the data description.
	This is in contrast with the difficulties  of the microscopic simulations mentioned above.
	The simplicity and powerfullness of our description reminds us the simplicity of the use of the ideal gas law,
	$pV = k NT$, to describe the expansion and related pressure, volume and temperature changes of an ideal gas,
	keeping the essential physical variables, instead of modelling such an expansion by Monte-Carlo simulations of the same process.
	If indeed relativistic hydrodynamics is relevant to describe the scaling properties of high energy proton-proton
	and heavy ion collisions, similar simplifications and data collapsing behaviour can be expected. 
	In this paper, we have given three
	specific examples for scaling laws of hydrodynamical origin: 
	the rapidity density of Eq.~(\ref{e:dndy-Gauss}), the rapidity dependence of the effective temperature 
	in Eq.~(\ref{e:mtbar-y}) and the pseudorapidity density of Eq.~(\ref{e:dndeta-function}).
	Each suggests  such data collapsing behaviour. These new scaling laws
	are now rigorously derived -- under certain conditions --
	from the CKCJ exact solutions~\cite{Csorgo:2018pxh,Csorgo:2018fbz} of relativistic hydrodynamics.
	It is time to start their more detailed experimental scrutinity in various symmetric collisions of high energy particle
	and nuclear physics.
	
	Fig.~\ref{fig:dndeta_xexe} indicates, that our  fit describes the data in a larger pseudo-rapidity domain than derived in Refs.~\citen{Csorgo:2018pxh,Csorgo:2018fbz}, thus one may conjecture that perhaps the result of the derivation is more general, than the derivation itself. 
	The search for possible  generalizations of our  derivation of the hydrodynamical scaling laws of the $dN/dy$ and $dN/d\eta_p$  distributions is one of our currently open research questions. 
	
	As a possible experimental test of the relevance of the CKCJ solutions in describing high energy heavy ion collisions,
	we have evaluated the CKCJ prediction for the not yet measured rapidity density for
	Xe+Xe data at $\sqrt{s_{NN}} = $ 5.44 TeV, in the 0-80 \% centrality class, using the same set of parameters,
	that we have obtained from the pseudorapidity distributions. We have indicated our prediction
	for $dN/dy$ on the same Fig.~\ref{fig:dndeta_xexe}, as the pseudorapidity density, 
	in order  to highlight the similarities and the differences between these two, kinematically different but physically similar  distributions.
	
	\subsection{Lifetime estimates for Au+Au collisions at RHIC}
\begin{figure}[h!] 
		\includegraphics[scale=0.425]{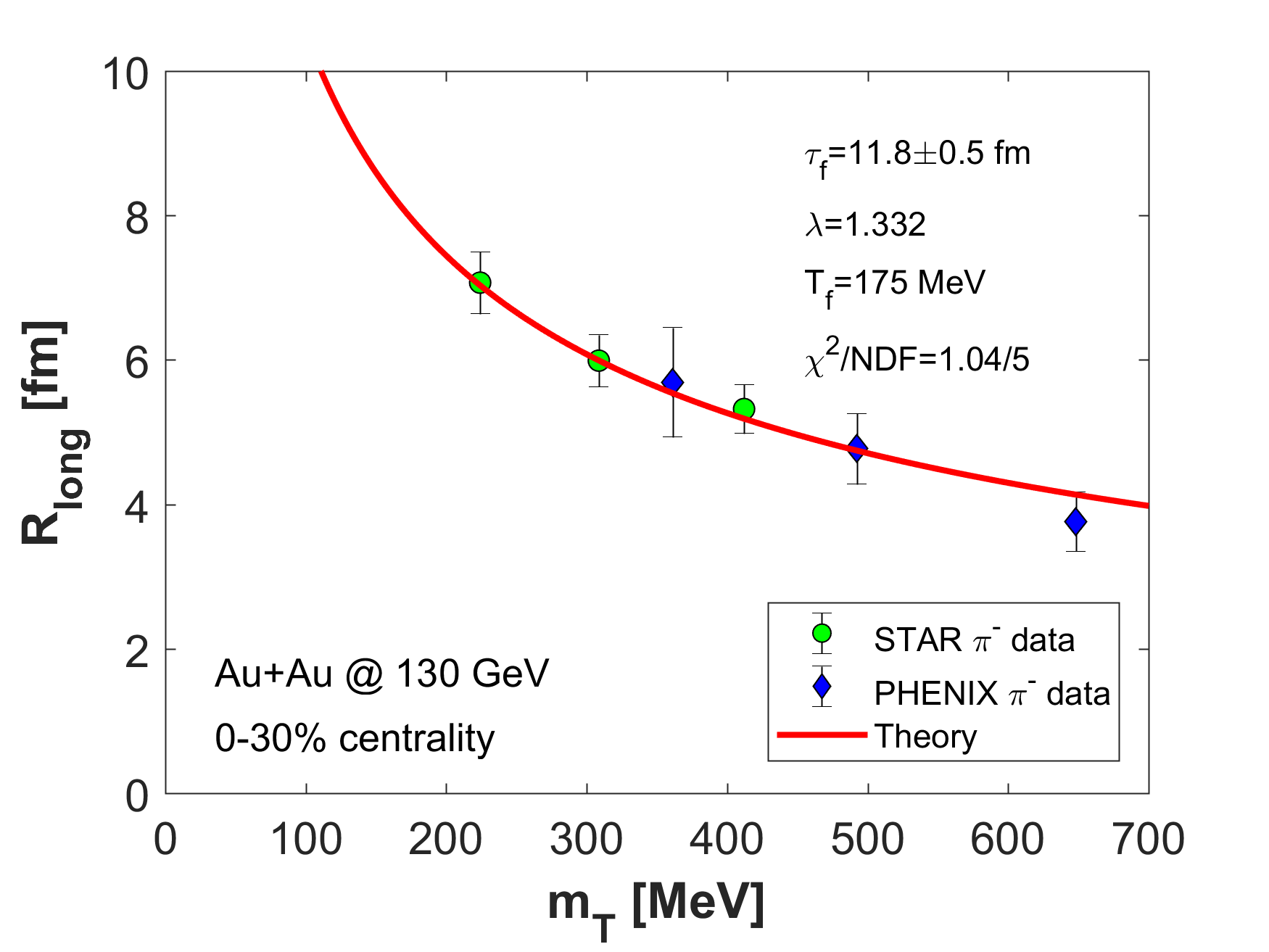}
		\includegraphics[scale=0.425]{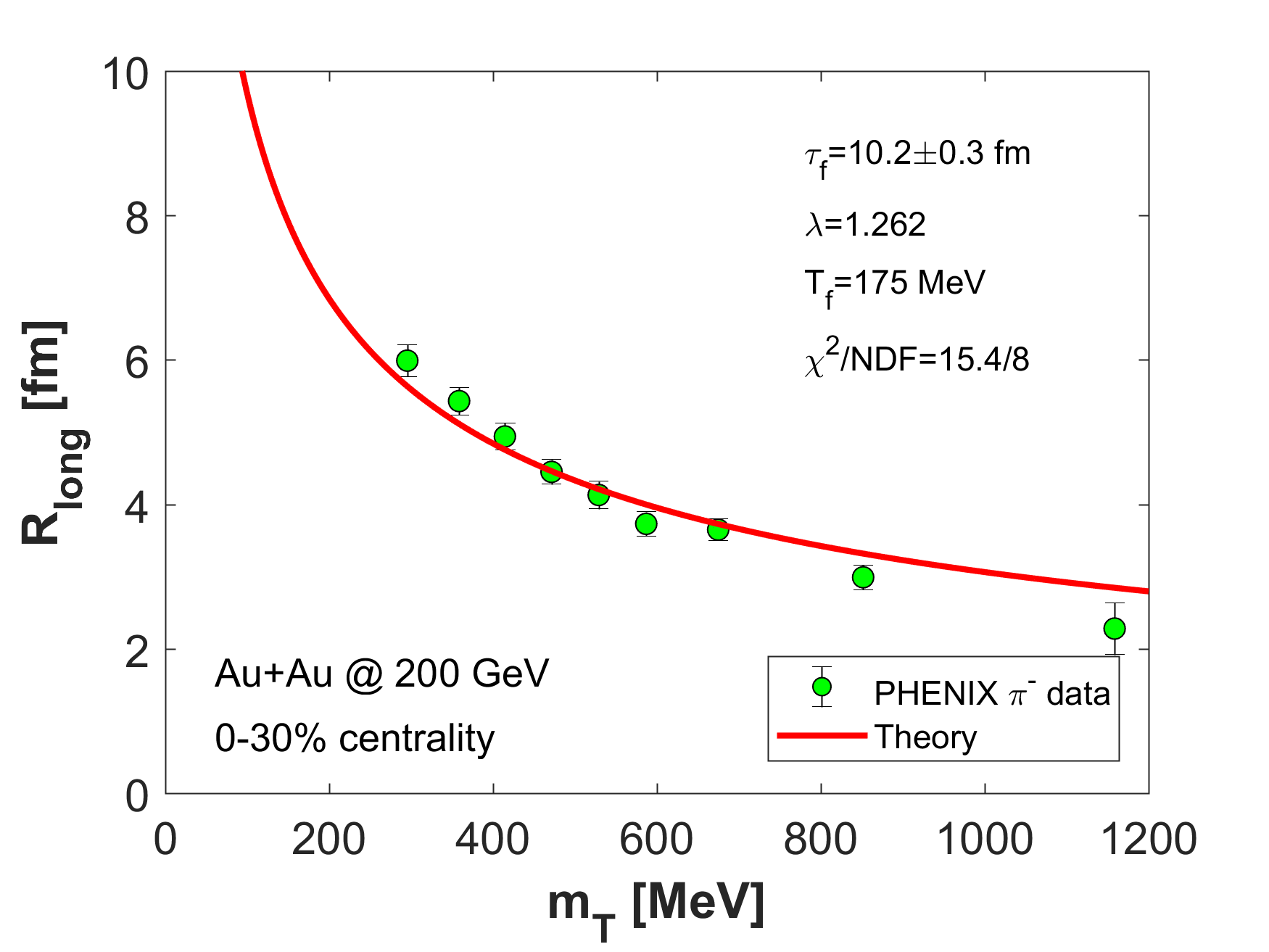}
		\centering 
		\caption{Fits of the longitudinal HBT-radii with the CKCJ hydro solution~\cite{Csorgo:2018pxh}, to PHENIX and STAR Au+Au data at $\sqrt{s_{NN}} = $ 130 GeV~\cite{Adcox:2002uc} (left) and to PHENIX Au+Au data at $\sqrt{s_{NN}} = $ 200 GeV~\cite{Adler:2004rq} (right) in the 0-30 \% centrality class, for a fixed centrality class and colliding system.}
		\label{fig:longitudinal-radius} 
	\end{figure}

	In order to extract the lifetime information of the fireball from Eq.~(\ref{e:Rlong-ckcj}), we compared the 
	life-time dependent longitudinal HBT radius parameters to experimental data.
	Our fits to the measured values of the longitudinal HBT radii $R_{\rm long}$  are illustrated on Figure~\ref{fig:longitudinal-radius}.
	The results are summarized in  Table~\ref{t:freezeout}, where 
	the systematic variation of the freeze-out time or life-time parameter $\tau_f$ with the variation of the assumed value of the freeze-out temperature $T_f$ is also indicated. Given that longer life-times yield larger initial energy density estimates for the same
	pseudorapidity distribution, we evaulated the initial energy densities for the more conservative values i.e. for shorter lifetimes, corresponding to
	$T_f = 175 $ MeV, which value is also consistent with the systematics of the transverse momentum spectra as indicated on Figure~\ref{fig:inverse_slope}.
	
	\begin{table}[tb]	
		\tbl{Freeze-out proper-times $\tau_f$ extracted from the transverse mass dependent longitudinal HBT radii using Eq.~\eqref{e:Rlong-ckcj} of the CKCJ exact solutions, assuming different values for the freeze-out temperature $T_f$.}		
		{\begin{tabular}{@{}ccc|cc@{}} \toprule
				$\sqrt{s_{NN}}$ & \multicolumn{2}{|c|}{Au+Au at 130 GeV,  0-30 \%}  & \multicolumn{2}{c}{Au+Au at 200 GeV, 0-30 \%}\\ \colrule
				\multicolumn{1}{c|}{$T_f$ [MeV]} & 140 & 175 & 140 & 175 \\
				\multicolumn{1}{c|}{$\tau_f$ [fm/c]} & 13.2 $\pm$ 0.6 & 11.8 $\pm$ 0.5 & 11.3 $\pm $ 0.4  & 10.2 $\pm$ 0.3 \\ \botrule
			\end{tabular}	 \label{t:freezeout}}
	\end{table} 
	
	With the help of these values of the lifetime parameter $\tau_f$ we are ready to calculate   the initial energy density of the expanding medium as a function of its initial proper time, $\tau_0$. This consideration serves the topic of the next section.
	
	\section{Initial energy density}
	
	In Ref.~\citen{Kasza:2018jtu} we have already shown an exact calculation of the initial energy density of the expanding fireball by utilizing the CKCJ hydrodynamic solution. We have found that Bjorken's famous estimate \cite{Bjorken:1982qr}, denoted by $\varepsilon_0^{\rm Bj}$, is corrected by the following formula, if the rapidity distribution is not flat and if the CKCJ solution is a valid approximation for the longitudinal dynamics:
	\begin{equation}\label{eq:IED-CKCJresult}
	\varepsilon_0(\kappa,\lambda)=\varepsilon_0^{\rm Bj} \left(2\lambda-1\right)\left(\frac{\tau_f}{\tau_0}\right)^{\lambda\left(1+\frac{1}{\kappa}\right)-1},
	\end{equation}
	where  Bjorken's (under)estimate for the initial energy density reads as
	\begin{equation}\label{eq:IED-bjorken}
	\varepsilon_0^{\rm Bj}=\frac{\langle E_T \rangle}{S_\perp \tau_0}\left.\frac{dN}{d\eta_p}\right|_{\eta_p=0}.
	\end{equation}
	In Bjorken's estimate $S_\perp$ is the overlap area of the colliding nuclei, and $\langle E_T \rangle$ is the average of the total transverse energy. In Eq.~\eqref{eq:IED-CKCJresult} the $\left(2\lambda-1\right)\left(\tau_f/\tau_0\right)^{\lambda-1}$ factor takes into account the shift of the saddle-point corresponding to the point of maximum emittivity for particles with a given rapidity,  and it takes into account also the change of the volume element during the non-boost-invariant expansion. 
	In addition to that, we have found an unexpected and in retrospective rather surprising feature of the energy density estimate from the CKCJ solution, as detailed in Ref.~\citen{Kasza:2018jtu}. Due to the EoS dependent correction factor, in the boost-invariant  limit, corresponding to $\lambda = 1$ and to the lack of acceleration, our estimate does {\it not} reproduce the  Bjorken estimate for the initial energy density. However, in this boost-invariant limit the CKCJ result for the initial energy density corresponds to Sinyukov's result from Ref.~\citen{Gorenstein:1977xv}.
	
	This difference between Bjorken's and our estimate for the initial energy density, that exists even in the boost invariant $\lambda =1 $ limit,  is related to the work, which is done by the pressure during the expansion stage.  If the pressure is vanishing,
	$p= 0$, it corresponds to the  equation of state of  dust, 
	and is obtained in the  $1/\kappa \rightarrow 0$ limit of our calculations. If
	$\lambda \rightarrow 1$ and $1/\kappa\rightarrow 0$, our estimate for the initial energy density reproduces Bjorken's estimate~\cite{Kasza:2018jtu}.
	In summary, the Bjorken estimate of the initial energy density has to be corrected not only due to the lack of boost-invariance
	in the measured  rapidity distributions, but also because it neglects that fraction of the  initial energy, 
	which is converted to the work done by the pressure during the expansion of the volume element in the center of the fireball~\cite{Kasza:2018jtu}. Bjorken was actually aware of this $p=0$ approximation in his original paper~\cite{Bjorken:1982qr}, which is a reasonable approximation  if 
	$p \ll \varepsilon$ and if one
	aims to get an order of magnitude estimations, that is  precise within a factor of 10. However, at the present age of complex and very expensive accelerators and  experiments, an order of magnitude or factor of 10 increase in initial energy density is less, than what can be gained by moving from the RHIC collider top energy of
	$\sqrt{s_{NN}} = 200$ GeV for Au+Au collisions to the $\sqrt{s_{NN}} = 2.76$ TeV for Pb+Pb collisions at LHC.
	Although the rise in the center of mass energy is more than a factor of 13 when going from the RHIC energy of $\sqrt{s_{NN}} = 200$ 
	GeV to the LHC energy of $\sqrt{s_{NN}} = 2.76$ TeV, the increase in the initial energy density is of the order of 2 only,  as demonstrated recently for example in Ref.~\citen{Ze-Fang:2017ppe} and summarized in Table~\ref{tab:zefang_IED}. 
	Note that the  values of $\kappa$-dependent estimates of the initial energy density,
	$\varepsilon^{\rm conj}(\kappa,\lambda)$ values  of Ref.~\citen{Ze-Fang:2017ppe} were based
	on a conjectured $\kappa$ dependence of the initial energy density, given in Ref.~\citen{Csorgo:2008pe} as follows:
	\begin{eqnarray}
	\varepsilon_0^{\rm CNC}(\lambda) & =  & \varepsilon_0^{\rm Bj} \left(2\lambda-1\right)\left(\frac{\tau_f}{\tau_0}\right)^{\lambda-1}, \label{e:epsilon0-CNC}\\ 
	\varepsilon_0^{\rm conj}(\kappa,\lambda) & =  & \varepsilon_0^{\rm CNC}(\lambda)
	\left(\frac{\tau_f}{\tau_0}\right)^{ (\lambda-1)\left(1-\frac{1}{\kappa}\right)}.\label{e:epsilon0-conj}    
	\end{eqnarray}
	The first of the above equations is an exact, proper-time dependent result for $\varepsilon_0^{\rm CNC}(\lambda)$,
	the initial energy density in the CNC solution of Refs.~\citen{Csorgo:2006ax,Nagy:2007xn,Csorgo:2008pe}.
	However, this Eq.~(\ref{e:epsilon0-CNC}) lacks the dependence of the initial energy density on the equation of state parameter $\kappa$.
	Nevertheless it contains two $\lambda$-dependent prefactors as compared to the Bjorken estimate, and so it allows for the possibility
	of a non-monotonic dependence of the initial energy density on the energy of the collision, if the shape and the transverse energy density
	both change monotonously but non-trivially with $\sqrt{s_{NN}}$. 
	
	The second estimate of the initial energy density, Eq.~(\ref{e:epsilon0-conj}) was a  conjecture, assumed at times when the CKCJ solution
	was not yet known. This conjecture  was obtained under the condition that it reproduces
	the exact CNC result in the $\kappa \rightarrow 1$ as well as in the $\lambda \rightarrow 1$ limits~\cite{Csorgo:2008pe}. 
	As compared to the CNC estimate, this conjecture contains a  proper-time dependent prefactor,
	that generalized the CNC estimate under the requirement that it has to reproduce the Bjorken estimate in
	the $\lambda \rightarrow  1 $ limit for all values of the parameter of the Equation of State $\kappa$.  This is reflected by the vanishing of the exponent of the second proper-time dependent factor in the $\lambda \rightarrow 1$ limiting case. However, by now we know that this requirement
	is not valid as the initial energy density has a so far largely ignored but also equation of state dependent prefactor.
	This is due to an actual $\kappa$ dependent (but neglected) term in the initial energy density in the boost-invariant $\lambda \rightarrow 1$ limiting case. Fortunately the conjecture was numerically reasonably good, as it contained the
	dominant and fastly rising $(2 \lambda -1) (\tau_f/\tau_0)^{\lambda-1}$ prefactor already,
	that increases with increasing values of $\lambda$ that corresponds to decreasing energies and increasing deviations from an asymptotic boost-invarance. Table~\ref{tab:zefang_IED}
	summarizes the Bjorken, the CNC exact and conjectured initial energy densities, and the CKCJ exact initial densities for $\sqrt{s_{NN}} =  $ 130 and 200 GeV, 0-30\% Au+Au collisions in the first two lines. The last line of the  same table also indicates 
	the Bjorken, the CNC and the CNC conjectured result for the initial energy density for 
	$\sqrt{s_{NN}} = 2.76$ TeV Pb+Pb collisions at CERN LHC, assuming the value of $\tau_f/\tau_0 = 10$. The first column of this Table~\ref{tab:zefang_IED} indicates that the Bjorken initial energy density increases monotonically with increasing $\sqrt{s_{NN}}$.
	The second column indicates that the acceleration parameter $\lambda$ dependent correction terms generate a {\it non-monotonic}
	energy dependence for the initial energy density already in the CNC model fit results. As the evolution of the shape
	of the pseudorapidity distribution is rather dramatic at lower energies, we can identify that the shape parameter $\lambda$ dependent terms in Eq.~(\ref{e:epsilon0-CNC}) create this {\it non-monotonic} behaviour, 
	that in turn is inherited by the more-and-more refined calculations,
	corresponding to the conjecture and to the CKCJ solution indicated in the last two columns of the same Table~\ref{tab:zefang_IED}. 
	
	Given that a more precise evaluation of the life-time parameter of $2.76$ TeV Pb+Pb collisions needs 
	more detailed analysis of this reaction with the help of the CKCJ solution,
	that goes beyond the scope of our already extended manuscript, the last cell in Table~\ref{tab:zefang_IED} is left empty, 
	to be filled in by future calculations.
	
	\begin{table}[tb]	
		\tbl{Initial energy density estimation from Ref.~\citen{Ze-Fang:2017ppe} by Bjorken's formula $\varepsilon_{0}^{\rm Bj}$, and the  conjectured values  of $\varepsilon_0^{\rm conj}(\kappa,\lambda)$, evaluated for $\tau_0 = 1 $ fm/c . These values also indicate the non-monotonic behaviour of the initial energy density as a function of $\sqrt{s_{NN}}$, the center of mass energy of colliding nucleon pairs.}		
		{\begin{tabular}{@{}ccccc@{}} \toprule
				& $\varepsilon_{0}^{\rm Bj}$ [GeV/fm$^3$] & $\varepsilon_0^{\rm CNC}(\lambda)$ [GeV/fm$^3$] & $\varepsilon_0^{\rm conj}(\kappa,\lambda)$ [GeV/fm$^3$] & $\varepsilon_0(\kappa,\lambda)$ [GeV/fm$^3$]\\ \colrule
				Au+Au at 130 GeV, 6-15 \% & 4.1 $\pm$ 0.4 & 14.8 $\pm$ 2.2 & 11.2 $\pm$ 1.8 & 11.9 $\pm$ 0.5  \\
				Au+Au at 200 GeV, 6-15 \% & 4.7 $\pm$ 0.5 & 12.2 $\pm$ 2.3 & 9.9 $\pm$ 1.6 & 9.8 $\pm$ 0.4 \\
				Pb+Pb at 2.76 TeV, 10-20 \% & 10.1 $\pm$ 0.3 & 14.1 $\pm$ 0.5 & 13.3 $\pm$ 0.6 &   \\ \botrule
			\end{tabular}	 \label{tab:zefang_IED}}
	\end{table}

\begin{figure}[h!] 
		\includegraphics[scale=0.425]{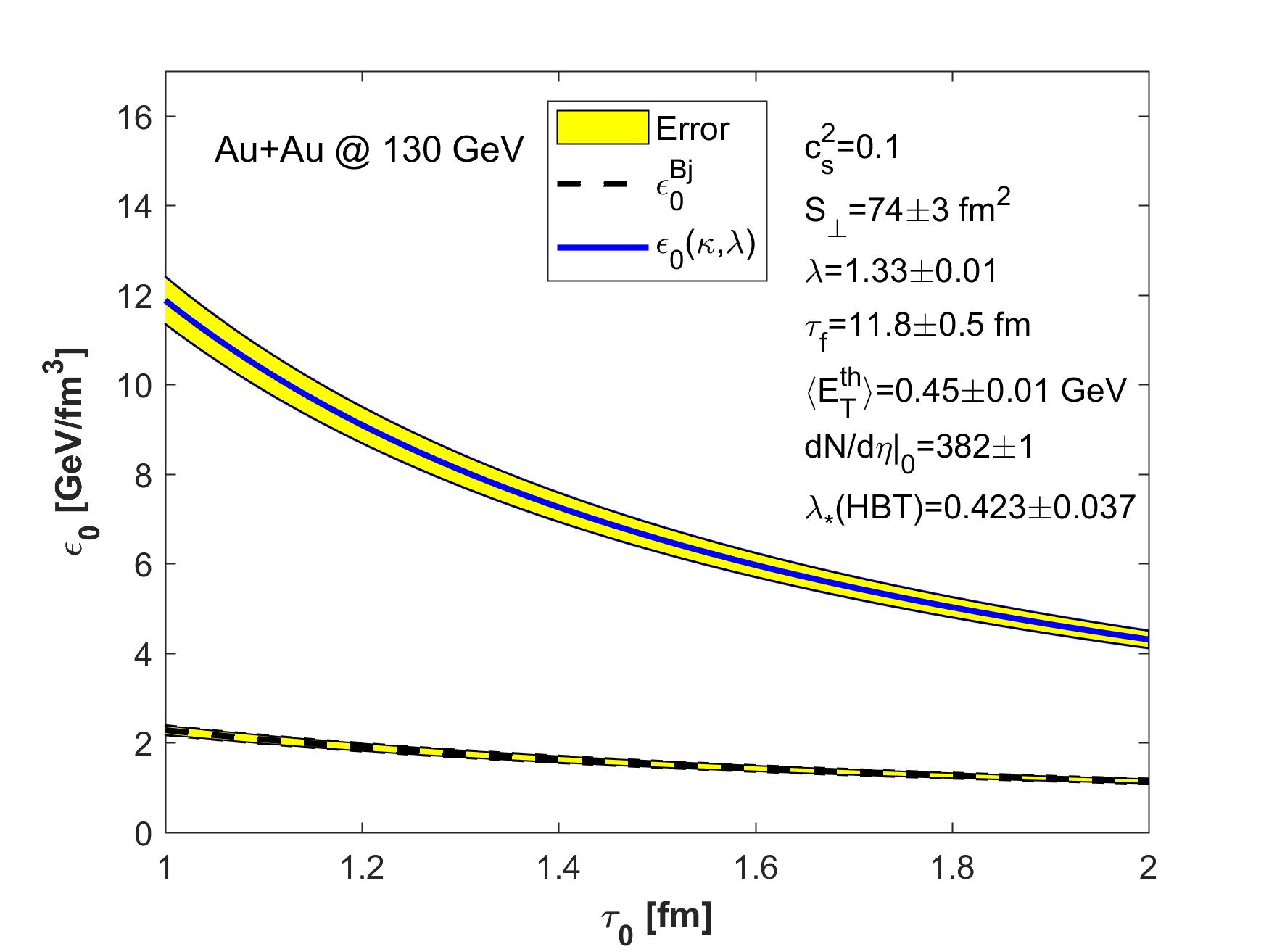}
		\includegraphics[scale=0.425]{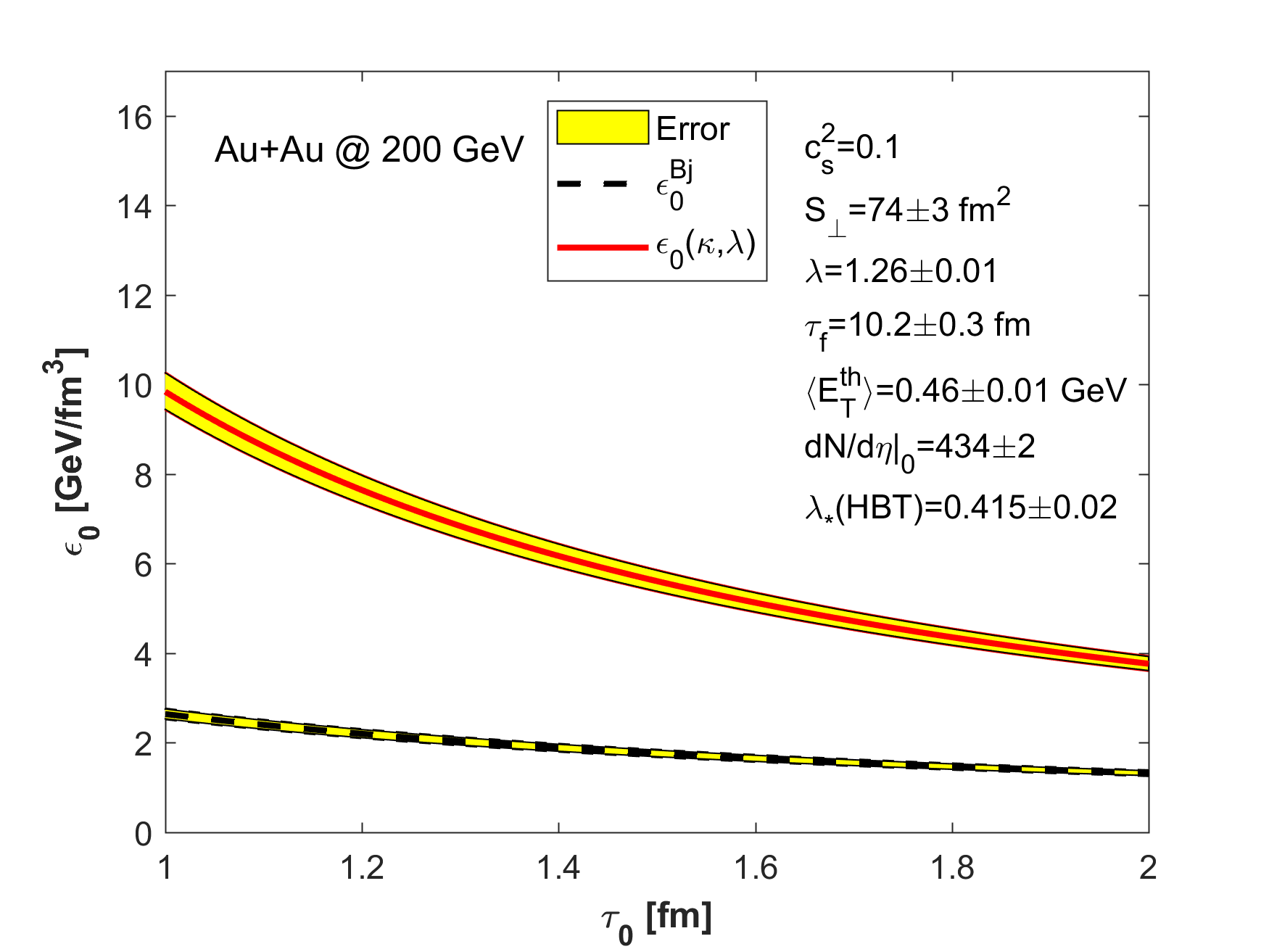}
		\centering 
		\caption{
			Initial energy density estimates from the CKCJ solution are shown with solid lines and compared to Bjorken's estimate, indicated with dashed lines, as a function of the initial proper time. The parameters of the left panel correspond to fit results of the CKCJ solution to PHOBOS Au+Au 
			pseudorapidity density data in the 0-30 \% centrality class both at $\sqrt{s_{NN}} = 130$ GeV
			(left panel)  at $\sqrt{s_{NN}} = 200$ GeV (right panel).}
		\label{fig:ied} 
	\end{figure}

\begin{figure}[h!] 
		\includegraphics[scale=0.425]{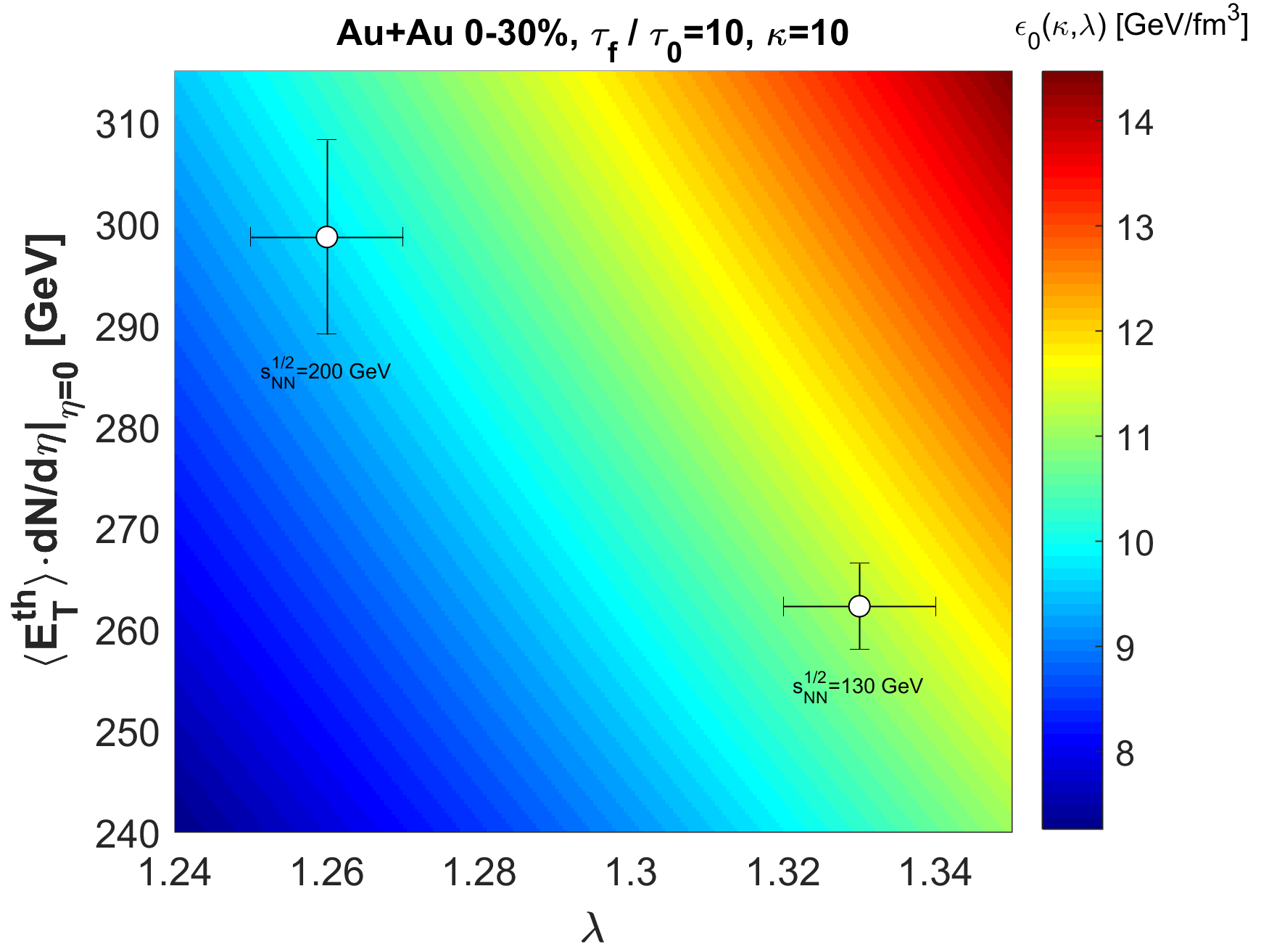}
		\includegraphics[scale=0.425]{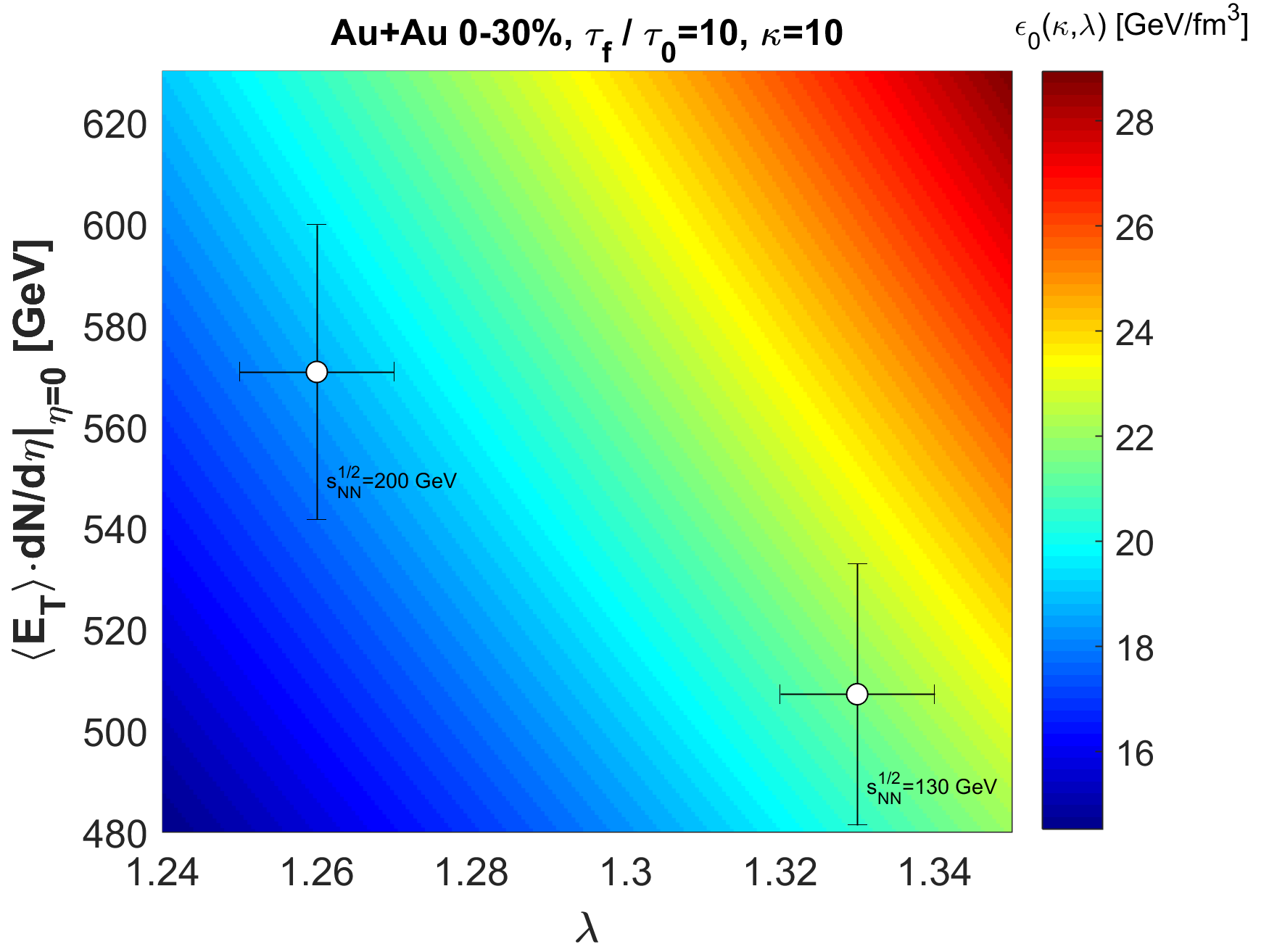}
		\centering 
		\caption{
			Location of the 0-30 \% central Au+Au collisions on the $(\lambda, dE_T/d\eta_p)$ diagram for $\sqrt{s} = 130$ and $200$ GeV.
			The color code indicates the contours of constant initial energy densities, evaluated for the realistic $\kappa = 10$ equation of state, corresponding to the measured speed of sound $c_s^2 \simeq 0.1$ in these reactions. Conservatively, these contours are evaluated for
			a $\tau_f/\tau_0 = 10$ ratio of the final over initial proper-time. The left panel indicates the thermalized energy density while the right panel indicates the pseudorapidity density of all transverse energy, including energy in non-thermal, high transverse momentum processes.
		}
		\label{fig:IED-map} 
	\end{figure}

	Let us use the fit parameters determined in Section~\ref{s:4} to the evaluation of our new, exact result for the initial energy density and for a comparison of it with Bjorken's estimate for the initial energy density, as a function of the initial proper time $\tau_0$. In Figure~\ref{fig:ied} we compare Eq.~\eqref{eq:IED-CKCJresult} to $\varepsilon_0^{\rm Bj}$ in  Au+Au $\sqrt{s_{NN}} = 130$ GeV and $\sqrt{s_{NN}} = 200$ GeV collisions. The normalization at midrapidity $\left.\frac{dN}{dy}\right|_{y = 0}$, the acceleration parameter $\lambda$ as well as the effective temperature $T_{\rm eff}$ are determined by fits shown on Figures~\ref{fig:inverse_slope} and \ref{fig:pseudorapidity-density}, while the value of the lifetime parameter $\tau_f$ is taken from fits illustrated in Figure~\ref{fig:longitudinal-radius}. The normalization parameter $\left.\frac{dN}{dy}\right|_{y = 0}$ corresponds to all the final state particles, and many of them are emitted from the decays of resonances, while the hydrodynamical evolution is restricted to describe the production of directly emitted particles that mix with
	the short-lived resonance decays whose life-time is comparable to the 1-2 fm/c duration of the freeze-out process. 
	This thermal part of hadron production can be calculated from the exact solutions of relativistic hydrodynamics, but they should be  corrected
	for the contribution of the decays of the long-lived resonances. According to the Core-Halo model \cite{Csorgo:1994in}, particle emission is divided into two parts. The Core corresponds to the direct production, which includes the hydrodynamic evolution and the short-lived resonances that decay on the time-scale of the freeze-out process. This part is responsible for the hydrodynamical behavior of the HBT radii and for that of the slopes of the single-particle spectra, a behaviour that is successfully describing the data shown in Figures~\ref{fig:inverse_slope},~\ref{fig:longitudinal-radius} as well as the shape of the pseudo-rapidity distributions on Fig.~\ref{fig:pseudorapidity-density}. On the other hand, the halo part consists of those particles, predominantly pions, that are emitted from the decays of long lived resonances. Exact hydrodynamic solutions,  such as the CKCJ solutions can make predictions only for the time evolution of the core. According to the core-halo model of Ref.~\citen{Csorgo:1994in} the normalization parameter of the pseudo-rapidity density, $\left.\frac{dN}{d\eta_p}\right|_{\eta_p = 0}$ can be corrected by a measurable core-halo correction factor, to get the contribution of the core
	from  Bose-Einstein correlation measurements:
	\begin{equation}
	\left.\frac{dN_{\rm core}}{d\eta_p}\right|_{\eta_p = 0} = \sqrt{\lambda_*^{\rm HBT}}\left.\frac{dN}{d\eta_p}\right|_{\eta_p =0},
	\end{equation}
	where the newly introduced the intercept of the two-pion Bose-Einstein correlation function, $\lambda_*^{\rm HBT}$  is taken from measurements
	of Bose-Einstein correlation functions.
	In principle this correction factor is transverse mass and pseudorapidity dependent. However, its transverse mass dependence is averaged out in the pseudorapidity distributions so we take its typical value at the average transverse mass, at midrapidity.
	In addition, the pseudorapidity dependence of this intercept parameter is, as far as we are aware of, not determined in
	Au+Au collisions at RHIC, given that the STAR and the PHENIX measurements of the Bose-Einstein correlation functions were performed at mid-rapidity,
	as both detectors identify pions around mid-rapidity only. In the core-halo model, its value is given as $\lambda_* ^{\rm HBT}=N_{\rm core}^2/N^2$ and its average value is measured through the HBT or Bose-Einstein correlation functions  for Au+Au $\sqrt{s_{NN}}$=130 GeV (Ref.~\citen{Adcox:2002uc}) and $\sqrt{s_{NN}}$=200 GeV (Ref.~\citen{Adler:2004rq}) as well. The size of the overlap area of the colliding nuclei,  $S_\perp$ 
	is estimated by the Glauber calculations of Ref.~\citen{Miller:2007ri}, and the average thermalized transverse energy
	is estimated from the nearly exponential shape  of the transverse momentum spectra in $m_T -m$ as follows:
	\begin{equation}
	\langle E_T^{th} \rangle=\left(m+T_{\rm eff}\right)\left(1+\frac{T_{\rm eff}^2}{\left(m+T_{\rm eff}\right)^2}\right).
	\end{equation}
	This estimate includes only the thermalized energy, which estimate corresponds to the CKCJ  solution estimate after being embedded into 1+3 dimensions. Correspondingly, the energy density present in high $p_T$ processes is  neglected by this formula and its applications yield a lower limit of the initial energy density. The measured values of the transverse energy production at mid-rapidity, however, include the
	effects of high transverse momentum processes from perturbative QCD. The difference between the two kind of initial energy densities,
	the thermalized source and the full initial energy density is illustrated as the difference between  the left and the right panels of Figure~\ref{fig:IED-map}.
	
	\begin{figure}[h!] 
		\includegraphics[scale=0.425]{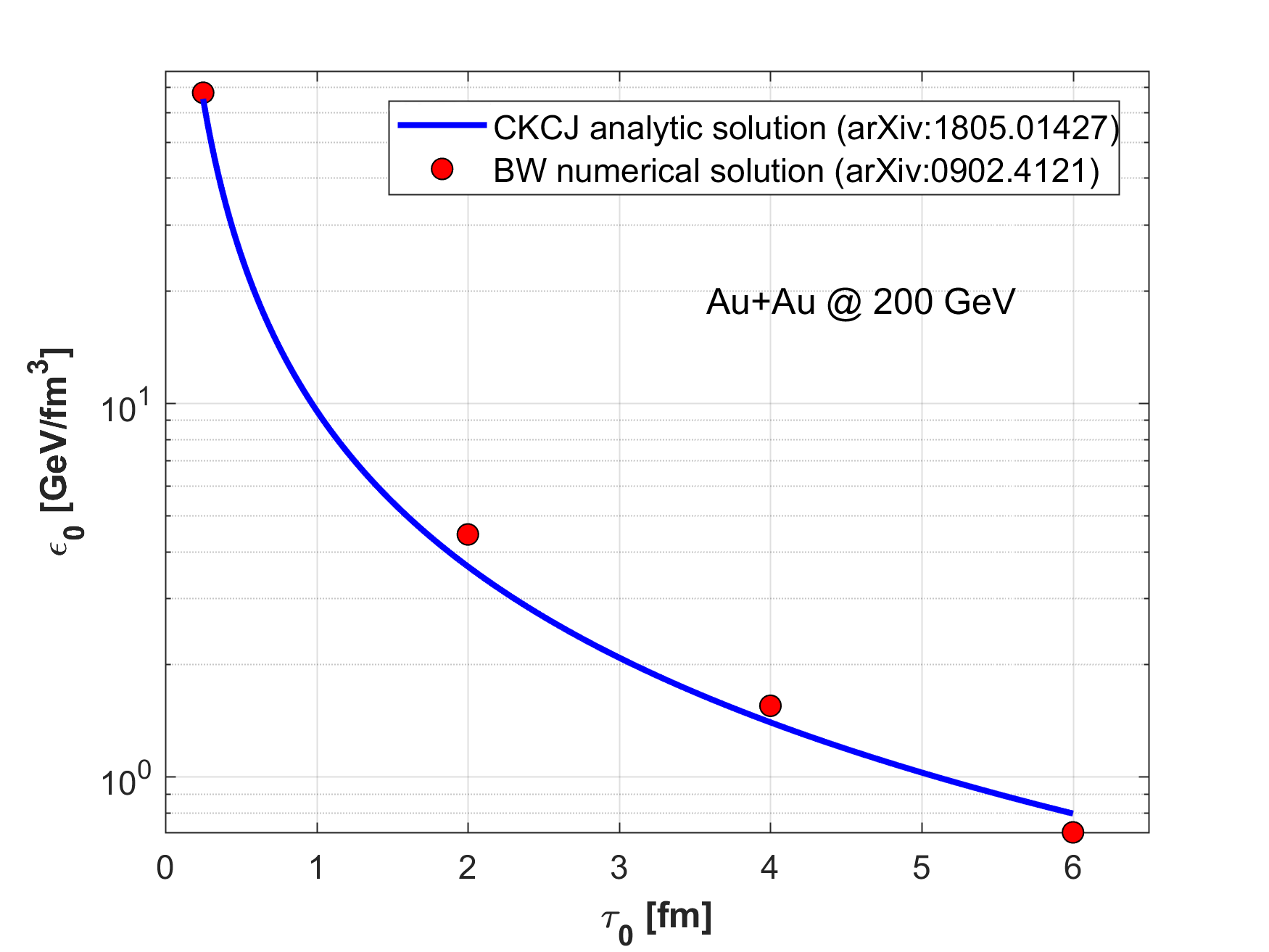}
		\includegraphics[scale=0.425]{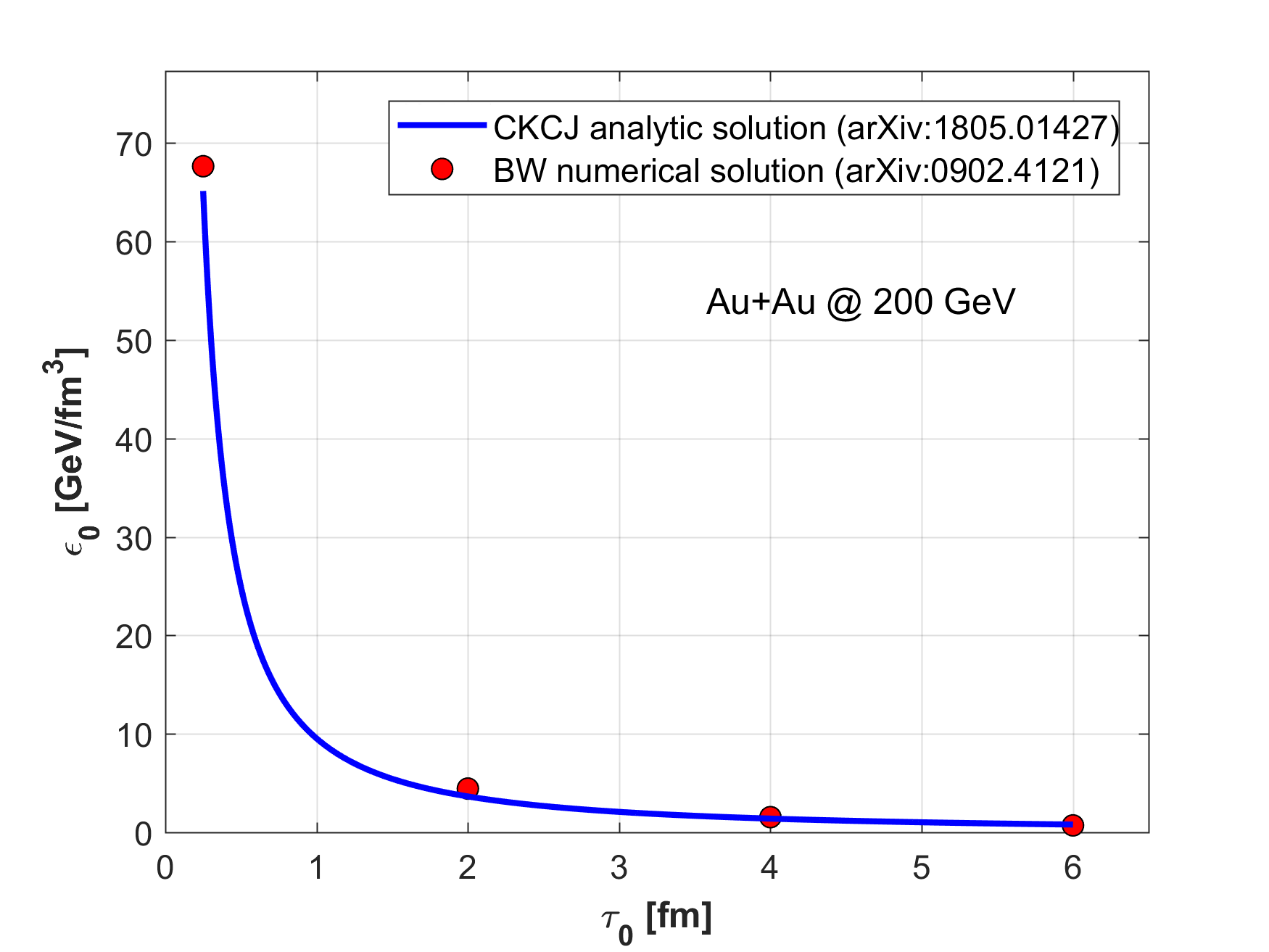}
		\centering 
		\caption{
			Initial energy density estimates from CKCJ analytic solution are show with blue solid lines and compared to the initial energy estimate from a numerical calculation of Bozek and Wyskiel~\cite{Bozek:2009ty}, indicated with red markers, as a function of the initial proper time. These (blue) solid curves correspond to the (red) solid  curves of the right panel Fig.~\ref{fig:ied} . The left panel shows the same plot as the right panel, but the left panel  uses a  logarithmic vertical scale for a comparison of the (red) markers with the (blue) solid line, while the right panel uses a linear vertical scale to show the same comparison.}
		\label{fig:ied_compare} 
	\end{figure}
	Figure~\ref{fig:ied} indicates, that Bjorken's formula underpredicts the initial energy density for both collision energies and that it predicts higher initial energy densities for higher collision energies, because the pseudorapidity density of the average transverse energy,
	$d\langle E_T\rangle/d\eta_p$ is a monotonously increasing function of the colliding energy $\sqrt{s_{NN}}$ for a given centrality class. 
	However, an unexpected and  really surprising feature is also visible on these plots: namely, our formula finds that   for the lower collision energies  of $\sqrt{s_{NN}} = 130$ GeV, the initial energy densities are  higher, due to the larger acceleration and work effects, than at the larger colliding energies of $\sqrt{s_{NN}}$ = 200 GeV! This unexpected behaviour is caused by 
	an interplay of two different effects. Although the midrapidity density is increasing monotonically with increasing colliding energy,
	that would increase the initial energy densities with increasing energy of the collision,
	the acceleration and the related work effects decrease with increasing energies. Although both $\frac{dE}{d\eta_p}|_{\eta_p=0}$ and $\lambda$ change as a monotonic function of $\sqrt{s_{NN}}$, $\varepsilon_0$ is a non-trivial function of both $\lambda$ and $\frac{dE}{d\eta_p}|_{\eta_p=0}$, so that
	the net effect is a decrease of the initial energy density with increasing colliding energies.
	Thus the CKCJ corrections of the Bjorken estimation have more significant effects in lower collision energy because of the greater acceleration and longer lifetime of the fireball. 
	
	This feature is clearly illustrated on both the left and the right panels of Fig.~\ref{fig:IED-map}. 
	The projection of the two data points to the vertical axis indicates that the
	transverse energy production is a monotonically increasing function with increasing $\sqrt{s_{NN}}$ for Au+Au collisions in the same
	centrality class. This feature is independent of the usage of the thermalized energy density (left panel) or the total available
	energy density (right panel). The projection of the two points to the horizontal axis is also inidicating a monotonously changing/flattening  shape of the (pseudo)rapidity distribution 
	with increasing $\sqrt{s_{NN}}.$ However, the contours of constant initial density are non-trivially dependent on both the vertical and the horizontal axes, and the location of the two points on the two-dimensional plot with respect to the contour lines indicates, that the initial energy density in $\sqrt{s_{NN}} = 200$ GeV Au+Au collisions is actually less than the initial energy density in the same reactions but at lower, 130 GeV colliding energies, in the same 0-30 \%
	centrality class.
	
	The astute reader may think at this point that such a non-monotonic behaviour of the initial energy density as a function of the 
	colliding energy may be due just to the 1+1 dimensional nature of the adobted CKCJ solution i.e. the neglect of the effects of transverse expansion
	on the transverse energy density and also due to the neglect of a possible temperature dependence of the speed of sound.
	
	To cross-check and clarify these questions, in Fig.~\ref{fig:ied_compare} we compared the CKCJ initial energy density estimate of Au+Au collisions at $\sqrt{s_{NN}}$=200 GeV in the 0-30\% centrality class to the 1+3 dimensional numerical solution of the equations of relativistic hydrodynamics by Bozek and Wyskiel (BW)~\cite{Bozek:2009ty}. This numerical result was shown to reproduce the pseudorapidity density of charged particles
	in $\sqrt{s_{NN}} = 200$ GeV Au+Au collisions, using lattice QCD equation of state, with a temperature dependent speed of sound. Both of the analytic and numerical estimations are corrected for the decays of (long-lived) resonances, so Fig.~\ref{fig:ied_compare} compares only the thermalized initial energy densities from the CKCJ analytic solution and from the BW numerical solution. The estimation of the analytic but 1+1 dimensional CKCJ solution is surprisingly similar to the numerical but 1 + 3 dimensional calculations as far as initial energy densities are concerned. The similarity of the evolution of the energy density in the center of the fireball in these two different hydrodynamical solutions
	may just be  a coincidence, given that Ref.~\citen{Bozek:2009ty} did not detail the centrality dependence of the time evolution of the energy density. However, the centrality dependence is expected to influence the overall normalization predominantly, so the similar 
	time-dependence of the decrease   of the energy density in the CKCJ solution and in the calculations of Ref.~\citen{Bozek:2009ty} 
	clearly indicates, that the collective dynamics in the center of the fireball is predominantly longitudinal and that transverse flow effects do not substantially modify the pseudorapidity density of transverse energy at mid-rapidity -- as conjectured by Bjorken in Ref.~\citen{Bjorken:1982qr}.
	
	Although the centrality classes and the estimations of the EoS dependence of the initial energy densities are somewhat different in the current work as compared to that of Ref.~\citen{Ze-Fang:2017ppe}, that paper provides an additional possibility for a cross-check.
	The results are summarized in Table~\ref{tab:zefang_IED}.
	It is important to realize, that in Ref.~\citen{Ze-Fang:2017ppe}  the critical energy densities as well as their non-monotonic dependence on $\sqrt{s_{NN}}$ are within errors the same as in the exact calculations presented in this paper, that can be seen from a comparison of the last two columns of Table~\ref{tab:zefang_IED}, where the  results in the last column correspond to Fig.~\ref{fig:ied}.

	\section{Summary}
	In this manuscript, we started to evaluate the excitation functions of the main characteristics of high energy heavy ion collisions
	in the RHIC energy range. The initial energy density and the lifetime of these reactions was estimated with the help of a novel family of exact and analytic, finite and accelerating, 1+1 dimensional solutions of relativistic perfect fluid hydrodynamics,  found recently by Cs\"org\H{o}, Kasza, Csan\'ad and Jiang~\cite{Csorgo:2018pxh}. With this new solution we evaluated the rapidity and the pseudorapidity densities and demonstrated, that these results describe well the pseudorapidity densities of Au+Au collisions at $\sqrt{s_{NN}} = 130$ GeV and $\sqrt{s_{NN}} = 200$ GeV in the 0-30\% centrality class. 
	
	We have obtained three remarkable theoretical results:
	\begin{itemize}
		\item We have derived a simple and beautiful formula to describe the pseudorapidity density distribution from the CKCJ solution of relativistic hydrodynamics~\cite{Csorgo:2018pxh}. This formula is given by Eqs.~\eqref{e:y-eta} and \eqref{e:dndeta-function}, and it describes not only $p+p$ and heavy ion data at RHIC and LHC energies, but it also describes the recent CMS data on Xe + Xe collisions exceedingly well.
		\item We have derived the scaling relation that at mid-rapidity,  $R_{\rm long } = A \big(\frac{dN}{d\eta_p}\big)^{1/3}$ follows from the CKCJ solution of relativistic hydrodynamics~\cite{Csorgo:2018pxh} .
		This scaling relation has been discovered empirically in Ref.~\citen{Lisa:2005dd} and this relation has been found to be valid for a broad set of data in a recent paper by the ALICE collaboration in Ref.~\citen{Adam:2015pya}, but, as far as we know, this relation
		has not been derived from relativistic hydrodynamics before.
		\item We have found a new method to extract the initial energy density of high energy proton-proton  and heavy ion collisions,
		that corrects Bjorken's initial energy density estimate for realistic pseudo-rapidity density distributions.
	\end{itemize}

	From fits to these pseudorapidity distributions, we determined the acceleration parameter $\lambda$ as well as the effective temperature, the slope of the transverse momentum distribution $T_{\rm eff}$. With the help of these values, we reduced the number of free fit parameters to one in the 
	analytic expression from the CKCJ solution that describes the transverse mass dependence of the longitudinal HBT-radii. We have determined the freeze-out proper time $\tau_f$ from transverse mass dependent fits to PHENIX and STAR $R_{\rm long}$ measurements. We have found that $\tau_f=11.8\pm0.5$ fm at $\sqrt{s_{NN}} = 130$ GeV and $\tau_f=10.2\pm 0.3$ fm at $\sqrt{s_{NN}} = 200$ GeV Au+Au collisions in the 0-30 \% centrality class. These values were utilized  to evaluate the initial energy densities, as a function of the initial proper time, for both collision systems. We compared our new, exactly calculated formula to Bjorken's estimate and the results were quite surprising. The CKCJ solution and the
	corresponding data analysis finds higher initial energy densities in $\sqrt{s_{NN}} $ = 130 GeV Au+Au collisions in the 0-30 \% centrality class as compared to the same kind of collisions
	at larger, $\sqrt{s_{NN}} = 200 $ GeV colliding energies. 
	\begin{itemize}
		\item The applications of our results to Au + Au  data at RHIC and Pb+Pb collisions at LHC energies
		suggest,  that the initial energy density is a non-monotonic function of the colliding energy at the RHIC energy range of $\sqrt{s_{NN}} \le 200$ GeV, given that at the LHC energies the initial energy densities increase,
		as indicated on Table~\ref{tab:zefang_IED}.
	\end{itemize} 
	We find that   the observed decrease of the initial energy density from 
	$\sqrt{s_{NN}} = 130$ GeV to 200 GeV collisions is an indication of a non-monotonic behaviour, that has a limited range,
	as the initial energy density increases from $\sqrt{s_{NN}} = 200$ GeV to $\sqrt{s_{NN}} = 2.76$ TeV heavy ion collisions. 
	It is thus important and urgent to map out the
	dependence of the initial energy density of high energy heavy ion collisions on the colliding energy
	for all currently available data sets. 

    Let us comment on the counterintuitive behavior of the initial temperature. Its dropping with  center of mass energy increasing from
    $\sqrt{s}_{NN}$ growing from 130 to 200 GeV apparently contradicts basic QCD-related ideas on the physics of multiparticle production, as the gluon density is expected to increase monotonically with increasing colliding energy. However, near the QCD critical point, several physical quantities like the initial conditions as well as final state observables are expected to behave non-monotonically. The detailed evaluation of the excitation function of the initial energy density as proposed in our manuscript
    may thus also become a new tool to search for the critical point of the QCD phase diagram:  in the vicinity of this critical point, several quantities may behave in a non-monotonic manner, including life-time related observables. For example, the non-monotonic behaviour of the HBT-radii has been observed in the RHIC energy domain, pointing to a QCD critical point near $\mu_B \approx 95$ MeV, corresponding to   $\sqrt{s_{NN}} \approx 47.5$ GeV~\cite{Lacey:2014wqa}. Our data analysis related to the estimations of the initial energy density of
    $Au+Au$ collisions at RHIC supports independently the possibility of such a scenario, although more detailed studies at lower RHIC energies and a refined treatement of the forward and backward rapidity regions are needed to try to locate the QCD critical point with our method. Based on these arguments we may expect that the initial energy density increases monotonically with increasing colliding energy, then in a limited
    range of colliding energies it may decrease with further increase (eg. from $\sqrt{s}_{NN} \ge 47.5$ GeV up to 200 GeV). However, at higher colliding energies, where the gluons dominate totally  the initial energy density, the initial temperatures and energy densities  may become again monotonically increasing functions of the colliding energy.
	
	In order to refine our resolution of our ``femtoscope"  further, and in order to deepen our understanding, the generalizations of the CKCJ solution to the cases of {\it i)} a temperature dependent speed of sound, {\it ii)} a  one plus three dimensional expansion, {\it iii)}  viscous solutions with shear and bulk viscosity terms are being explored at the time of closing this manuscript.
	
	
	\section*{Acknowledgments}
	
	We would like to thank M. Csan\'ad,  L. L\"{o}nnblad, R. Pasechnik, G. Gustafson and J. Schukraft for clarifying, inspiring and  useful discussions. T. Cs. expresses his gratitude to R. Pasechnik for inviting the presentation of these results
	to the COST Workshop on the Interplay of Hard and Soft QCD Probes for Collectivity in Heavy Ion Collisions and for a kind hospitality at the University of Lund. It is our pleasure to thank one of the Referees of IJMPA for his or her insights and constructive criticism during the anonymous peer-review process. Our research has been partially supported  by the THOR project, COST Action CA15213 of the European Union,
	as well as by the Hungarian NKIFH grants FK-123842 and FK-123959 and the EFOP-3.6.1-16-2016-00001 grants (Hungary).

	
\end{document}